\documentclass[11pt,a4paper]{article}
\usepackage{jheppub}
\usepackage{amssymb}
\usepackage{subfig}

 \newlength{\wth}
\newcommand{\sixgraphs}[6]{%
\unitlength=1in
\begin{picture}(8.7,7.1)(-0.7,-0.1)
\put(-0.5,4.8){\epsfig{file=#1, width=\wth}}
\put(2.7,4.8){\epsfig{file=#2, width=\wth}}
\put(-0.5,2.3){\epsfig{file=#3, width=\wth}}
\put(2.7,2.3){\epsfig{file=#4, width=\wth}}
\put(-0.5,-0.2){\epsfig{file=#5, width=\wth}}
\put(2.7,-0.2){\epsfig{file=#6, width=\wth}}
\end{picture}}

\begin{document}
\title{Fast supersymmetry phenomenology at the Large Hadron Collider using machine learning techniques}
\author[a]{A. Buckley,}
\author[b]{A. Shilton,}
\author[c]{and M. J. White}
\affiliation[a]{University of Edinburgh, School of Physics \& Astronomy, Mayfield Rd, Edinburgh, EH9\,3JZ, UK}
\affiliation[b]{University of Melbourne, School of Engineering, Department of Electrical and Electronic Engineering, VIC 3010, Australia}
\affiliation[c]{University of Melbourne, School of Physics, VIC 3010, Australia}
\emailAdd{mwhi@unimelb.edu.au}

\abstract{A pressing problem for supersymmetry (SUSY) phenomenologists is how to incorporate Large Hadron Collider search results into parameter fits designed to measure or constrain the SUSY parameters. Owing to the computational expense of fully simulating lots of points in a generic SUSY space to aid the calculation of the likelihoods, the limits published by experimental collaborations are frequently interpreted in slices of reduced parameter spaces. For example, both ATLAS and CMS have presented results in the Constrained Minimal Supersymmetric Model (CMSSM) by fixing two of four parameters, and generating a coarse grid in the remaining two. We demonstrate that by generating a grid in the full space of the CMSSM, one can interpolate between the output of an LHC detector simulation using machine learning techniques, thus obtaining a superfast likelihood calculator for LHC-based SUSY parameter fits. We further investigate how much training data is required to obtain usable results, finding that approximately 2000 points are required in the CMSSM to get likelihood predictions to an accuracy of a few per cent. The techniques presented here provide a general approach for adding LHC event rate data to SUSY fitting algorithms, and can easily be used to explore other candidate physics models.}

\maketitle
\flushbottom
\section{Introduction}
The ATLAS and CMS experiments~\cite{Aad:2008zzm,cms:2008zzk} at the Large Hadron Collider at CERN, Geneva, have recently published their first results relating to the search for new theories of particle physics. Of the variety of theories developed to explain the deficiencies of the Standard Model, supersymmetry (SUSY) is a leading contender. The absence of evidence for supersymmetric particle production in the first LHC data can be used to derive limits on the parameters of candidate supersymmetric theories~\cite{Aad:2011ks,Aad:2011hh,daCosta:2011qk,Chatrchyan:2011wc,Khachatryan:2011tk,Dolan:2011ie,Buchmueller:2011aa,Buchmueller:2011ki,Bechtle:2011it,Akula:2011zq,Allanach:2011wi}. This in turn requires one to calculate the number of expected signal events at different points in the SUSY parameter space, before assigning a suitable likelihood based on the number of observed events and the Standard Model background expectation.

To rigorously estimate the number of expected signal events for a point in a SUSY parameter space, one must evaluate the cross-section and then perform a Monte Carlo simulation to obtain a realistic sample of `reconstructed' candidate events, with correctly modelled acceptance and resolution effects. The time for such a simulation ranges from several minutes for a fast, approximate simulation code to hours for the full GEANT4-based simulations employed by the experimental collaborations. A next-to-leading order evaluation of the total SUSY production cross-section might take a further half an hour or so. Clearly neither of these approaches can be used directly in a fit to supersymmetric parameters where one often requires millions of likelihood evaluations in order to obtain a reasonable result. The ATLAS and CMS collaborations therefore interpret their results in restricted model spaces such as the Constrained Minimal Supersymmetric Model (CMSSM), fixing all parameters except the unified scalar and gaugino mass parameters $m_0$ and $m_{1/2}$ (the remaining parameters being the trilinear coupling term $A_0$, the Higgs VEV ratio $\tan\beta$ and the sign of the Higgs mass parameter $\mu$). One can then fully simulate (i.e., simulate both the fully hadronised event and the detector interaction/reconstruction effects) a grid of points in parallel upon which to set an exclusion limit. This limit can be interpreted by phenomenologists, but is only really useful if the number of events in the search channels under study is only weakly dependent on $A_0$ and tan$\beta$, whence one can use the measurement in that channel to make more general inferences~\footnote{Of course, it is also useful in the case that nature has chosen the same values of $A_0$ and $\tan\beta$ as those found in the ATLAS and CMS papers, but we consider this unlikely.}.

This paper will suggest a more general solution to the problem, using as a test case the full 4 parameters of the CMSSM (we have continued to assume $\mu>0$). If one could successfully interpolate between the simulation output values in the full space of the CMSSM (or any other model space), one would obtain a function that could be evaluated in fractions of a second to give the expected number of signal events in a given search channel, and this could subsequently be used for very fast likelihood calculations. We demonstrate that this is possible using either a Bayesian neural net or a support vector machine as a regressor (in an approach borrowed from the cosmology literature~\cite{Auld:2006pm}), with training data supplied by a fast, public LHC simulation code of the kind routinely used in the phenomenology literature. Our results could therefore already be used to do LHC parameter fit studies in this context, whilst even the fast simulation would still have proved prohibitively expensive to use directly in a non-parallelised fitting routine that relied on, say, Markov Chain Monte Carlo techniques. We will also argue that this approach to adding LHC event rate data to SUSY parameter fitting algorithms solves many of the problems associated with previous attempts found in the literature, such as in References~\cite{Lester:2005je} and~\cite{Dreiner:2010gv}. If the LHC sees direct evidence of sparticle production in the near future, one could use an interpolation of inclusive variables to add extra information to SUSY parameter fits alongside other features such as kinematic endpoint information.

In addition, we investigate whether it is feasible for experimental collaborations to try a similar technique on the output of their full detector simulations, thus enabling them to provide a suitable likelihood function for phenomenologists after generating their model grids in the full CMSSM rather than in a 2D parameter slice. Although our work is carried out in the framework of the CMSSM, there should be no barrier to replicating these techniques in other models, though models with more parameters would inevitably require larger training data sets.

The paper is structured as follows. In Section~\ref{framework} we briefly review the machine learning techniques used in our analysis, covering both the Bayesian training of neural networks, and the use of support vector machines. Readers who do not need an introduction to these topics may safely skip this section. In Section~\ref{results} we give details of our training data simulation, and demonstrate the use of a neural net to obtain a fast CMSSM event rate calculator using the recent ATLAS zero lepton search as an example. We repeat the study using a support vector machine, and show the optimum results obtained for both algorithms when we did not restrict the size of our training data sample. Section~\ref{training} shows the performance for differing amounts of training data in order to examine the feasibility of using similar techniques with slower detector simulations than those utilised here. A detailed discussion of the implications of our results is deferred to Section~\ref{discussion}, where we briefly compare our work to previous attempts at the problem, clearly state the current limitations, and share further insight for those wishing to try similar techniques. Finally, we present conclusions in Section~\ref{conclusions}.

\section{Machine learning techniques for regression}
The problem we address in this paper can be stated as follows. Assume that we have calculated the number of signal events expected in an LHC detector for $n$ points in the CMSSM space, each defined by a given choice of $m_0$, $m_{1/2}$, $A_0$ and $\tan\beta$ (we have fixed $\mu$ to be positive, but could repeat our procedure for negative $\mu$). Is it possible to obtain an interpolating function for the output values, such that, for any new points that are not in our simulated sample, we can get the number of events expected in an LHC detector?

This is a standard regression problem, and there are a number of documented methods for addressing it.  A popular choice is a Multilayer Perceptron Network, which can be shown to be able to approximate any smooth function if the network has sufficient complexity.  Clearly the success in our example will depend on whether the target distribution of events in the CMSSM parameter space is indeed smooth. The quality of our results must serve to demonstrate whether this is true or not.  A popular alternative to Multilayer Perceptrons is Support Vector Regression, which applies the principles of structural risk minimization to achieve a trade-off between empirical risk minimisation and over-fitting.  We now present a brief introduction to both techniques.

\label{framework}
\subsection{Multilayer Perceptron Networks}
\subsubsection{Overview}
\begin{figure}
\begin{center}
  \includegraphics[width=0.58\textwidth]{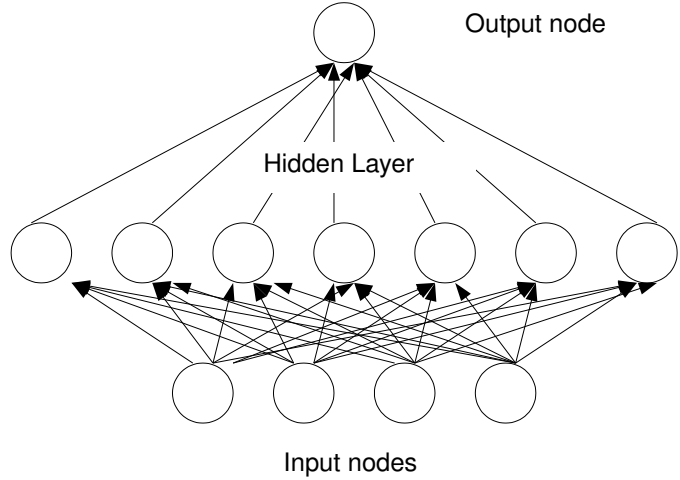}
\caption{Schematic diagram of a Multilayer Perceptron Network.}
\label{fig:net}
\end{center}
\end{figure}

The inspiration for the Multilayer Perceptron Network (MLP) is the architecture of animal brains. The networks consist of several `neurons', each of which sums a series of weighted scalar inputs, and produces an output determined by an `activation function' which is usually chosen to be either a step function or a non-linear function that varies between -1 and 1. The MLP network then consists of a series of linked neurons, as illustrated in Figure~\ref{fig:net}, and is an example of a `feed-forward' network in which information is passed through from inputs to outputs. For a detailed guide to neural networks and the Bayesian approach to network training, we refer the reader to~\cite{Neal:1996:BLN:525544}. For a convenient, short exposition with an example of application in a completely different context see~\cite{Vehtari20001183}.

In our case, the inputs are values of $m_0$, $m_{1/2}$, $A_0$ and $\tan\beta$, represented hereafter by the vector of input parameters $\textbf{p}$. The LHC target variable is the cross-section of events in a given search channel $y(\textbf{p})$ (we will assume that the process may be repeated for different search channels, and hence only have one output variable per net). The outputs of the output node and hidden nodes in the network might be calculated as follows:
\begin{equation}
f(\textbf{p},\textbf{w}) = b + \Sigma_j v_{j}h_j(x)
\end{equation}
\begin{equation}
h_j(x)=\tanh(a_j + \Sigma_i u_{ij}x_i)
\end{equation}
where $f(\textbf{p})$ gives the output of the output node (which in an ideal world would equal $y(\textbf{p})$), $u_{ij}$ is the weight on the connection from input unit $i$ to hidden unit $j$, $v_{j}$ is the weight on the connection from hidden unit $j$ to the output unit, and $b$ and $a_j$ are bias terms for the output and hidden nodes respectively. $\textbf{w}$ is used to label the entire set of network parameters for later convenience. Here we see that the output value is simply a weighted sum of the hidden unit values plus a bias term, whilst the hidden units put the weighted sum of their input values plus bias through a non-linear activation function. It is the non-linear nature of this activation function that allows the network to model complex distributions, and the tuning of the weights essentially controls the degree of complexity of the model.

To solve the problem of this paper, then, one needs to define a network architecture and find the values of the weights and biases of the neurons such that the network reproduces the correct value of the LHC target variable when fed the corresponding input data. This is accomplished by `training' the network with a set of training data $\{\textbf{p}^{(i)},y^{(i)}\}$. Several approaches to this problem exist, such as adjusting the weights and biases to minimise the sum of the squared differences between the network outputs $f(\textbf{p})^{(i)}$ and the correct values $y^{(i)}$. The function we wish to minimise may have many discrete minima, and hence ensuring that a correct solution is found is non-trivial.

\subsubsection{Bayesian approach to neural network training}
A particularly powerful approach to neural net training uses the methods of Bayesian inference. We first use our MLP network to define a probabilistic model for our regression problem as follows. Regression with a real valued target can be modelled by assuming that the target value is described by the output of our network, $f(\textbf{p},\textbf{w})$, subject to corruption from Gaussian noise of constant standard deviation $\sigma$. The probability density for the target is then:
\begin{equation}
\label{probdens}
p(y|\textbf{p},\textbf{w},\sigma)=\frac{1}{\sqrt{2\pi}\sigma}\text{exp}(-\frac{(y-f(\textbf{p},\textbf{w}))^2}{2\sigma^2})
\end{equation}

Let us call the full set of model parameters $\theta$, where $\theta \equiv \textbf{w},\sigma$ encompassing both the network parameters and the noise. Our prior knowledge of these parameters before any data have been observed can be represented by a prior distribution $p(\theta)$, and Bayes' theorem then tells us how to update the prior distribution to a posterior distribution after the training data $D=\{\textbf{p}^{(i)},y^{(i)}\}$ have been observed:
\begin{equation}
p(\theta|D)=\frac{p(D|\theta)p(\theta)}{p(D)}\propto L(\theta|D)p(\theta)
\end{equation}
Assuming that our training data points are independent, the likelihood $L(\theta|D)$ in this equation is given by a product of terms of the form of equation~\ref{probdens}, with the product running over all training points:
\begin{equation}
L(\theta|D)=\displaystyle\prod_{i=1}^np(y^{(i)}|\textbf{p}^{(i)},\theta)
\end{equation}
 If we now wish to obtain a new output $y^{(n+1)}$ from an input vector $\textbf{p}^{(n+1)}$ that is not in the training sample $D$, we must integrate the predictions of the model with respect to the posterior of the model parameters:
\begin{equation}
p(y^{(n+1)}|\textbf{p}^{(n+1)},D)=\int p(y^{(n+1)}|\textbf{p}^{(n+1)},\theta)p(\theta|D)\mathrm{d}\theta
\end{equation}
This defines the predictive distribution for $y^{(n+1)}$. If we want to estimate the value of this distribution that minimises the squared error loss, we can take the mean of this distribution:
\begin{equation}
\label{integral}
\hat{y}^{(n+1)}=\int f(\textbf{p}^{(n+1)},\theta)p(\theta|D)\mathrm{d}\theta
\end{equation}
\subsubsection{Implementation}
One can easily pick networks from the prior distribution on the network parameters as defined in our probabilistic model. The difficulty in the Bayesian approach to network training lies in evaluating the integral in equation~\ref{integral} to get our output value estimate. Since it is the expectation value of $f(\textbf{p}^{(n+1)},\theta)$ with respect to the posterior distribution of the parameters, however, one can use a Monte Carlo method using a sample of values $\theta^{(t)}$ drawn from the posterior distribution of parameters:
\begin{equation}
\hat{y}^{(n+1)}\approx \frac{1}{N}\displaystyle\sum\limits_{t=1}^n f(\textbf{p}^{(n+1)},\theta^{(t)})
\end{equation}
For our analysis, we use the Flexible Bayesian Modelling software (FBM) by Radford Neal~\footnote{\tt URL: http://www.cs.toronto.edu/\textasciitilde radford/fbm.software.html \rm}, which uses a combination of Markov Chain Monte Carlo (MCMC) methods to obtain these posterior samples and to evaluate equation~\ref{integral}. It assumes a Gaussian prior on the network weights, but allows the width of the Gaussian to be specified by a hyperparameter that is in turn specified by a vague prior. Hence, one essentially requires no intuitive knowledge of a suitable prior for the network parameters, but can instead let the data tune the complexity of the model. A vague prior is also used for the noise parameter $\sigma$. Neal uses the Hybrid Monte Carlo algorithm to explore the posterior of the network parameters, and periodically updates the hyperparameters using Gibbs sampling. This minimises the amount of tuning required to obtain good performance with the Hybrid Monte Carlo method.

\subsection{Support Vector Regression}
Support vector machines (SVMs)~\cite{Vap1,Bur1,Cri4} are a popular alternative to multi-layer perceptron type networks for both classification and regression problems~\cite{Wan6,Xia2,Mas1,Jay5,Lai4}.  Support vector regressors~\cite{Smo11,Vap4,Smo4} apply the principles of structural risk minimisation~\cite{Vap4} to achieve a trade-off between empirical risk (training set error) minimisation (ERM) and regularisation to avoid the problem of over-fitting.  They also have relatively fewer parameters to select in comparison with the MLP and do not suffer from the problem of local minima.

The support vector regression approach seeks to approximate the relation between input parameters ${\bf p} \in \mathbb{R}^{d_L}$ ($d_L$ being the number of parameters in the input) and targets $y$ using a trained machine of the form:
\[
 y \left( {\bf p} \right) = \left< {\bf w}, {\mbox{\boldmath $\varphi$}} \left( {\bf p} \right) \right> + b
\]
where ${\mbox{\boldmath $\varphi$}} : \mathbb{R}^{d_L} \to \mathbb{R}^{d_H}$ is the feature map into a $d_H$-dimensional {\em feature space} given a-priori, ${\bf w} \in \mathbb{R}^{d_H}$ the weight vector and $b \in \mathbb{R}$ the bias.  Given a training set of $N$ pairs $\left( {\bf p}^{(i)}, y^{(i)} \right)$ the weight vector ${\bf w}$ and bias $b$ are chosen to solve the primal training problem:
\begin{equation}
 \begin{array}{l}
  \begin{array}{l}
   \min\limits_{{\bf w}, b} R \left( {\bf w}, b \right) = \frac{1}{2} \left< {\bf w}, {\bf w} \right> + \frac{C}{N} \sum\limits_{i \in \mathbb{Z}_N} \xi_i
  \end{array} \\
  \begin{array}{ll}
   \mbox{such that:} & \left< {\bf w}, {\mbox{\boldmath $\varphi$}} \left( {\bf p}^{(i)} \right) \right> + b \leq y^{(i)} + \epsilon + \xi_i \; \forall i \in \mathbb{Z}_N \\
                     & \left< {\bf w}, {\mbox{\boldmath $\varphi$}} \left( {\bf p}^{(i)} \right) \right> + b \geq y^{(i)} - \epsilon - \xi_i \; \forall i \in \mathbb{Z}_N \\
                     & \xi_i \geq 0 \; \forall i \in \mathbb{Z}_N \\
  \end{array} \\
 \end{array}
\label{eq:svr_primal}
\end{equation}
where the second term in $R$, via the slack variables $\xi_i$ and together with the constraints, provides a measure of the training set error or empirical risk and the first term is a regularisation included to minimise over-fitting.  The training parameter $C \in \mathbb{R}^+$ controls the trade-off between these dual objectives, while $\epsilon \in \mathbb{R}^+$ controls noise insensitivity.

In practice the primal training problem of equation~\ref{eq:svr_primal} is rarely solved directly.  Instead the Wolfe-dual of equation~\ref{eq:svr_primal}:
\begin{equation}
 \begin{array}{l}
  \begin{array}{l}
   \min\limits_{{\mbox{\boldmath $\alpha$}}} Q \left( {\bf w}, b \right) = \frac{1}{2} \sum\limits_{i,j \in \mathbb{Z}_N} K_{ij} \alpha_i \alpha_j - \sum\limits_{i \in \mathbb{Z}_N} \alpha_i y^{(i)} + \epsilon \sum\limits_{i \in \mathbb{Z}_N} \left| \alpha_i \right|
  \end{array} \\
  \begin{array}{ll}
   \mbox{such that:} & -\frac{C}{N} \leq \alpha_i \leq \frac{C}{N} \; \forall i \in \mathbb{Z}_N \\
                     & \sum\limits_{i \in \mathbb{Z}_N} \alpha_i = 0 \\
  \end{array} \\
 \end{array}
\label{eq:svr_dual}
\end{equation}
is solved~\cite{Smo4}, where $K_{ij} = K \left( {\bf p}^{(i)}, {\bf p}^{(j)} \right)$, $K \left( {\bf p}, {\bf q} \right) = \left< {\mbox{\boldmath $\varphi$}} \left( {\bf p} \right), {\mbox{\boldmath $\varphi$}} \left( {\bf q} \right) \right>$ is the kernel function, and each dual variable $\alpha_i$, $i \in \mathbb{Z}_N$, corresponds to training pair $\left( {\bf p}^{(i)}, y^{(i)} \right)$.  Note that the dual in equation~\ref{eq:svr_dual} is a convex quadratic programming problem with a positive semi-definite Hessian ${\bf K} = \left[ K_{ij} \right]$, and so has no non-global minima.  The trained machine may also be written in terms of the dual variables $\alpha_i$:
\[
 y \left( {\bf p} \right) = \sum\limits_{i \in \mathbb{Z}_N} \alpha_i K \left( {\bf p}^{(i)}, {\bf p} \right) + b
\]

The advantage of the dual form over the primal form is that any function $K : \mathbb{R}^{d_L} \times \mathbb{R}^{d_L} \to \mathbb{R}$ satisfying Mercer's condition~\cite{Mer1,Tri1} may be used directly without explicit knowledge of the feature map ${\mbox{\boldmath $\varphi$}} : \mathbb{R}^{d_L} \to \mathbb{R}^{d_H}$ encompassed therein, allowing $d_H \gg d_L$ without added complexity.  This is known as the kernel trick.  Some popular kernel functions include~\cite{Her2,Ste3}:
\begin{enumerate}
 \item Linear kernel: $K \left( {\bf p}, {\bf q} \right) = \left< {\bf p}, {\bf q} \right>$.
 \item Polynomial kernel: $K \left( {\bf p}, {\bf q} \right) = \left( 1 + \left< {\bf p}, {\bf q} \right> \right)^d$ (where $d \in \mathbb{Z}^+$).
 \item RBF kernel: $K \left( {\bf p}, {\bf q} \right) = \exp \left( -\frac{1}{\sigma} \left\| {\bf p} - {\bf q} \right\|^2 \right)$ (where $\sigma \in \mathbb{R}^+$).
item Sigmoid kernel: $K \left( {\bf p}, {\bf q} \right) = \tanh \left( \kappa \left< {\bf p}, {\bf q} \right> + \beta \right)$ (where $\kappa, \beta \in \mathbb{R}^+$).
\end{enumerate}

For simulation purposes polynomial and RBF kernels have been used. Simulations have been carried out using SVMHeavy.\footnote{URL: http://people.eng.unimelb.edu.au/shiltona/svm/index.html}

\section{ATLAS signatures in the CMSSM}
\label{results}
\subsection{Signal region and simulation details}
In order to demonstrate the utility of machine learning methods in interpolating LHC simulation output, we will develop one example derived from the recent SUSY search results published by the ATLAS experiment. The most constraining search published at the time of writing is that performed in the zero lepton channel, which used four different signal regions to search for an excess of events over the Standard Model background. We will take as a test case `signal region D', defined by the cuts given in Table~\ref{cuts}, where $\Delta\phi($jet, $p_{T}^\text{miss})_\text{min}$ is the smallest of the azimuthal separations of $p_{T}^\text{miss}$ and the hardest jets with $p_T>40$ GeV (up to a maximum of three) and $m_\text{eff}$ is defined as the sum of $E_T^\text{miss}$ and the magnitudes of the transverse momenta of the three hardest jets. This is sufficient to prove the principle of machine-learning based interpolation, and we leave a detailed examination of all signal regions (including those yet to be published) for future work.
\begin{table}
\begin{center}
\begin{tabular}{|c|c|}
\hline
Signal region&D\\
\hline
Number of jets&$\geq 3$\\
Leading jet $p_{T}$[GeV]&$>120$\\
Other jet(s) $p_{T}$[GeV]&$>40$\\
$E_T^\text{miss}$[GeV]&$>100$\\
$\Delta\phi($jet, $p_{T}^\text{miss})_\text{min}$&$>0.4$\\
$E_T^\text{miss}/m_\text{eff}$&$>0.25$\\
$m_\text{eff}$[GeV]&$>1000$\\
\hline
\end{tabular}
\end{center}
\caption{\label{cuts} Event selection cuts used in the ATLAS experiment zero lepton search for supersymmetric particles~\cite{daCosta:2011qk}. See text for variable definitions.}
\end{table}

Training data for our support vector machine and neural net runs was generated by picking points at random from a specified distribution in the CMSSM parameter space then, for each point, running \tt Isajet 7.75\rm~\cite{Paige:2003mg} to generate the sparticle mass and decay spectra, followed by \tt HERWIG 6.505\rm~\cite{Corcella:2000bw,Corcella:2002jc,Moretti:2002eu} with \tt AcerDet\rm~\cite{RichterWas:2002ch} to generate a sample of 50,000 events. We then apply the selection cuts given in Table~\ref{cuts} and store the cross-section of events entering signal region D, $\sigma^{(\text{D})}$ (this is equivalent to storing the number of events observed in the search channel once the integrated luminosity is specified). The number of events generated and simulated per point must reduce the statistical error on $\sigma^{(\text{D})}$ to an acceptable level: 50,000 events is more than enough to negate the effect of statistical fluctuations in the training data on the quality of the interpolation results.

In choosing a range for the CMSSM parameters, we took care to ensure that our choice was reasonably large but, nevertheless, that it covered most of the interesting region for early LHC data. Our final choice is given in Table~\ref{prior1}. Generating data for 20,000 of these training points took approximately 1.5 days on the University of Melbourne Tier 2 cluster. It would take considerably longer with a full detector simulation, and thus there is a clear need to investigate how big the training data set needs to be to obtain reasonable results. This is considered in the next section.

\begin{figure}
\begin{center}
  \includegraphics[width=0.48\textwidth]{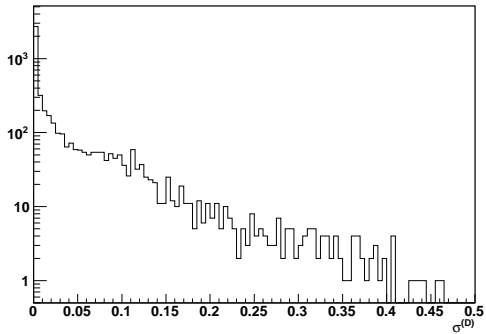}
  \caption{Distribution of the cross-section of events passing the selection cuts for region D (at leading order), for 5000 training points of 50,000 events each.}
\label{fig:yD}
\end{center}
\end{figure}

A histogram of the output values obtained for ATLAS Region D is given in Figure~\ref{fig:yD}, for CMSSM parameters taken from a flat distribution in the range specified in Table~\ref{prior1}. One can see that the true cross-section has a tail that is underpopulated by generating training data in this fashion. One can fix this is a variety of ways, of which the simplest options are:
\begin{enumerate}
\item Generate extra data in the tail of the distribution and train two nets to cover both regions. For CMSSM data, the points in the tail span the entire range of $m_0$, $A_0$ and $\tan\beta$, but have $m_{1/2}\lesssim 500$. We therefore tried two training sets of equal size: one in the full mass range, and one with $m_{1/2}<500$.
\item Generate training data from a distribution chosen so as to maximise the number of points in the tail. We did this by picking $m_0$ and $m_{1/2}$ values at random from the distribution $f(x;\lambda)=e^{-\frac{x}{\lambda}}/\lambda$, with $\lambda=200$, whilst retaining a flat distribution in  $A_0$ and $\tan\beta$.
\end{enumerate}
The two methods gave very similar results and hence we only show results for the second case. The ability to train one net instead of two, with a reduced total training set size makes the second option strongly preferred.

In addition to our training data set, we generated an independent test set of 5,000 events in order to present our final results, and to easily compare the support vector machine and Bayesian neural net methods. The support vector machine requires a training and test sample at the training stage, whilst the Bayesian neural net only requires a training sample. By using a final test set that is completely independent of any of this data, we ensure that we get a fair comparison between our two methods. It would not be necessary to generate such a large test set when applying the interpolation method in practice.
\subsection{Interpolation results}
\subsubsection{Results obtained using a Bayesian neural net}
We trained an MLP network with 2 hidden layers of 50 neurons each, and assessed convergence of the FBM software by looking at the variation of the squared error on the training sample vs iteration number, as well as the variations in the net weights and biases. Convergence was deemed to have set in once the squared error on the training set stabilised, and once the weights commenced a stable, periodic variation around a fixed value. In general, nets with less training data required more iterations before convergence was observed, with $\approx 1300$ iterations required for 5000 training points.
\begin{table}
\begin{center}
\begin{tabular}{|c|c|c|}
\hline
Parameter&Minimum&Maximum\\
\hline
$m_{1/2}$[GeV]&50&1200\\
$m_{0}$[GeV]&50&1200\\
$A_{0}$[GeV]&-1000&1000\\
$\tan\beta$&2&60\\
\hline
\end{tabular}
\end{center}
\caption{\label{prior1} Range used for the CMSSM parameters, in which we work to obtain an optimum interpolation between the output values of the LHC simulation. }
\end{table}

\begin{figure}
\begin{center}
  \subfloat[]{\label{fig:comp1}\includegraphics[width=0.48\textwidth]{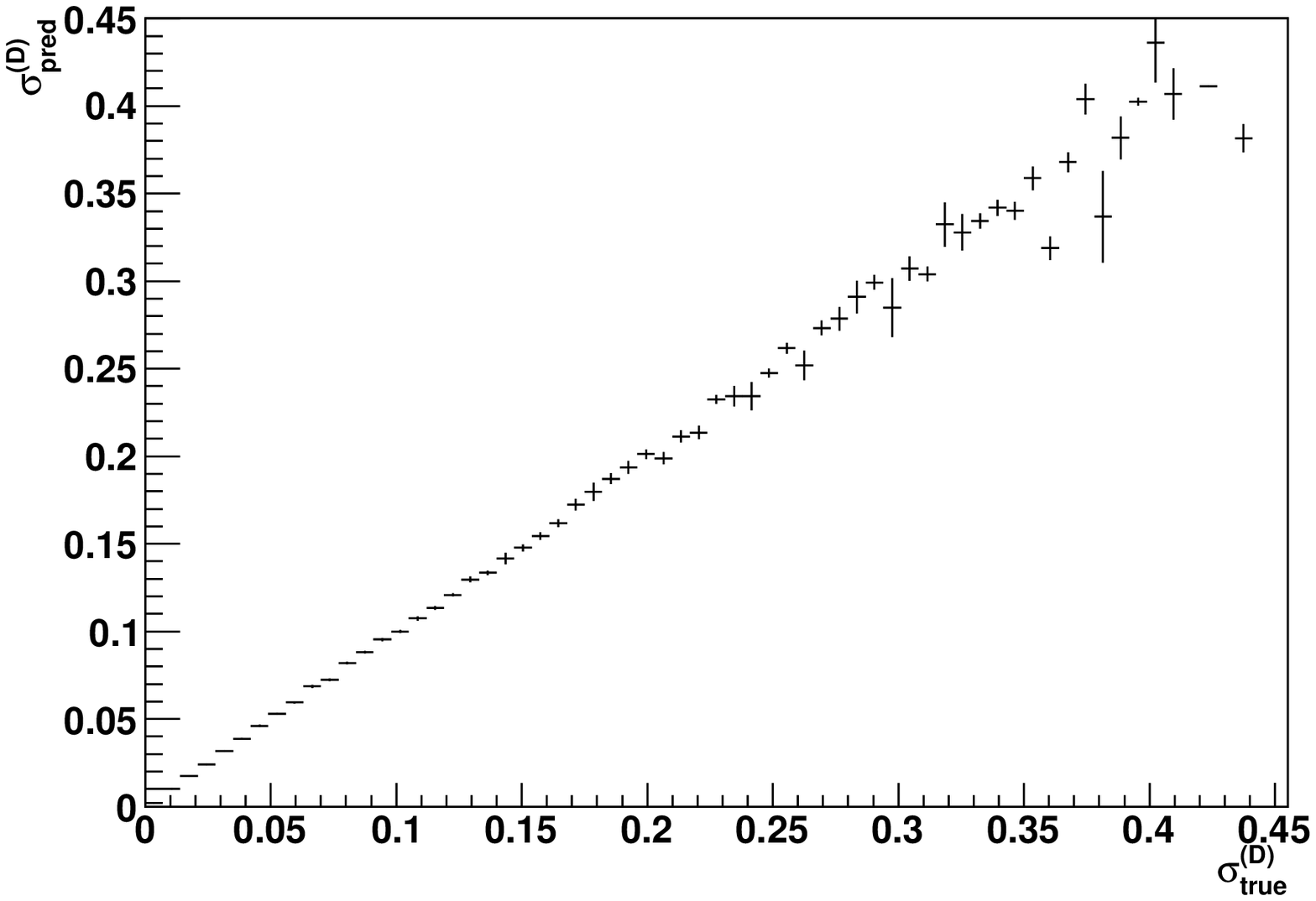}}
  \subfloat[]{\label{fig:comp2}\includegraphics[width=0.48\textwidth]{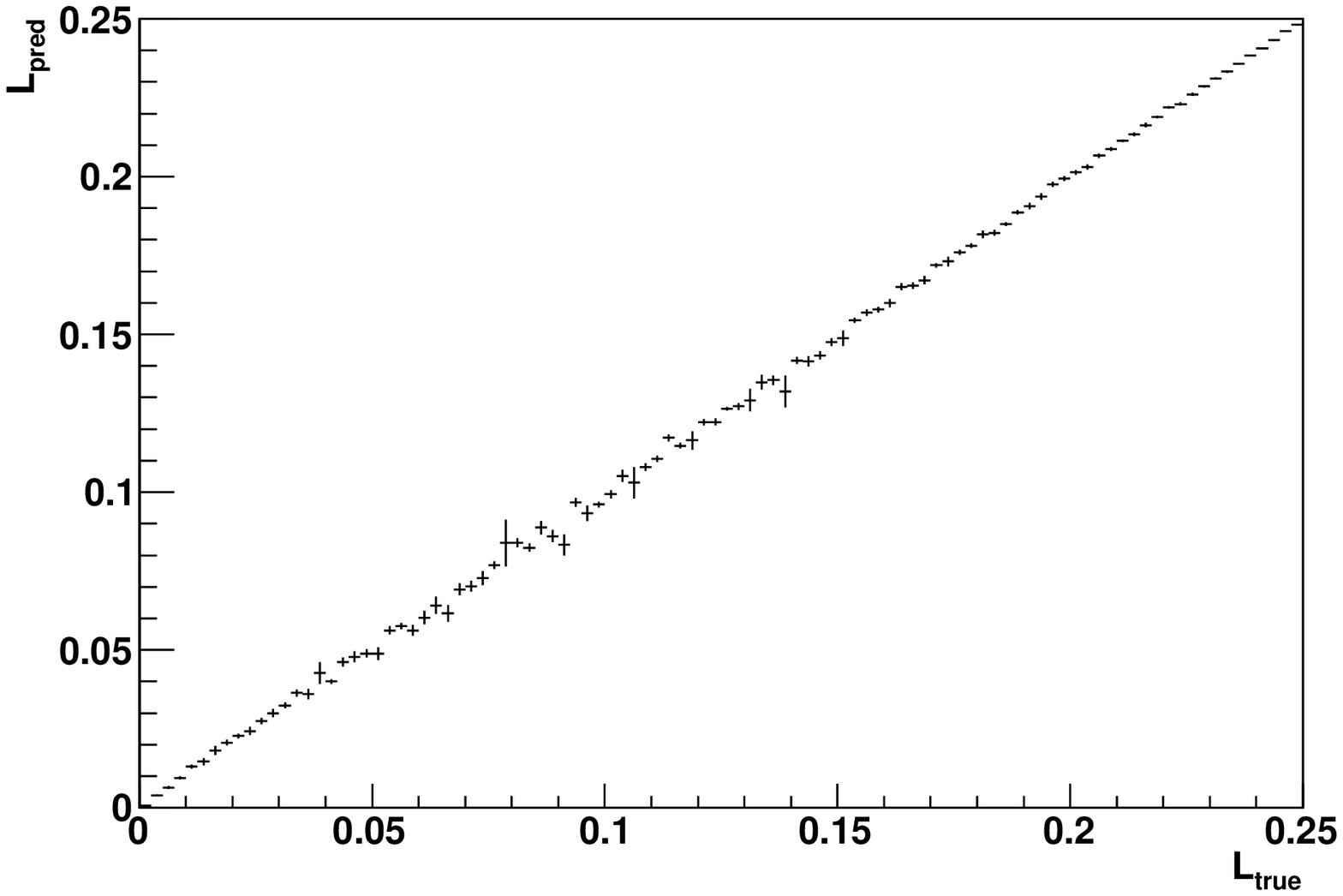}}
\caption{The true cross-section of events passing the ATLAS region D selection cuts (left), and the true Poisson likelihood of points in the CMSSM space (right) shown for the original simulation output vs. the prediction obtained using a Bayesian neural net trained with 5000 simulated data points and evaluated on an independent test sample of 5000 points.}
\label{fig:comp}
\end{center}
\end{figure}

Figure~\ref{fig:comp1} shows our predicted cross-section of events passing the ATLAS Region D selection cuts ($\sigma^{(\text{D})}_\text{pred}$) vs the true value from the full Monte Carlo simulation and reconstruction ($\sigma^{(\text{D})}_\text{true}$) as evaluated on our independent test sample, using a net trained with 5000 training points. The results demonstrate excellent agreement for low to moderate values of the true cross-section, with the performance degrading towards higher values. Here it is important to note that the error is still reasonable even in the tail, and also that this is the region in which one can tolerate a larger error. Since ATLAS has not seen a significant excess over the Standard Model cross-section, points with relatively large signal cross-sections are already very unlikely, and will not figure prominently in a parameter fit. To demonstrate this, one can evaluate the Poisson likelihood for observing $o$ events with a signal(background) expectation of $s(b)$:
\begin{equation}
L=\frac{e^{-(s+b)}(s+b)^o}{o!}
\end{equation}

Figure~\ref{fig:comp2} shows this likelihood for the values of $o$ and $b$ published in Reference~\cite{daCosta:2011qk}, where we neglect the effects of statistical and systematic uncertainties for this illustration. Figure~\ref{fig:comp2} demonstrates that our net output reproduces the ATLAS likelihood within a few percent over the entire range of interest, thus supporting our conclusion that it is not essential to model the entire tail of the signal cross-section distribution to per cent accuracy in order to obtain excellent final results. Indeed, the selection of training data is partly an exercise in using physical intuition to decide where in the CMSSM to expend most effort on detailed simulation.

\begin{figure}
  \subfloat[]{\label{fig:limit1}\includegraphics[width=0.48\textwidth]{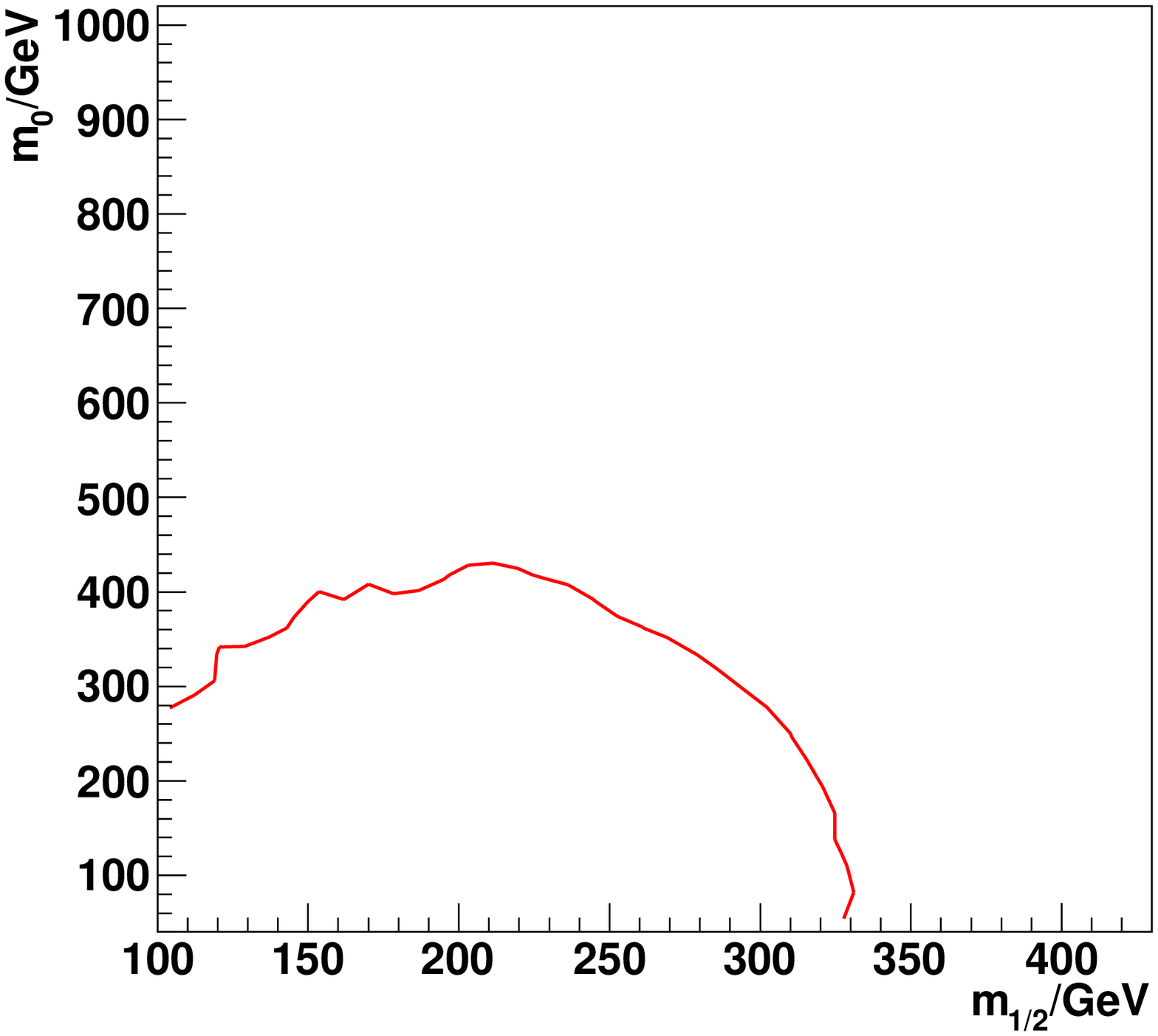}}
  \subfloat[]{\label{fig:limit2}\includegraphics[width=0.48\textwidth]{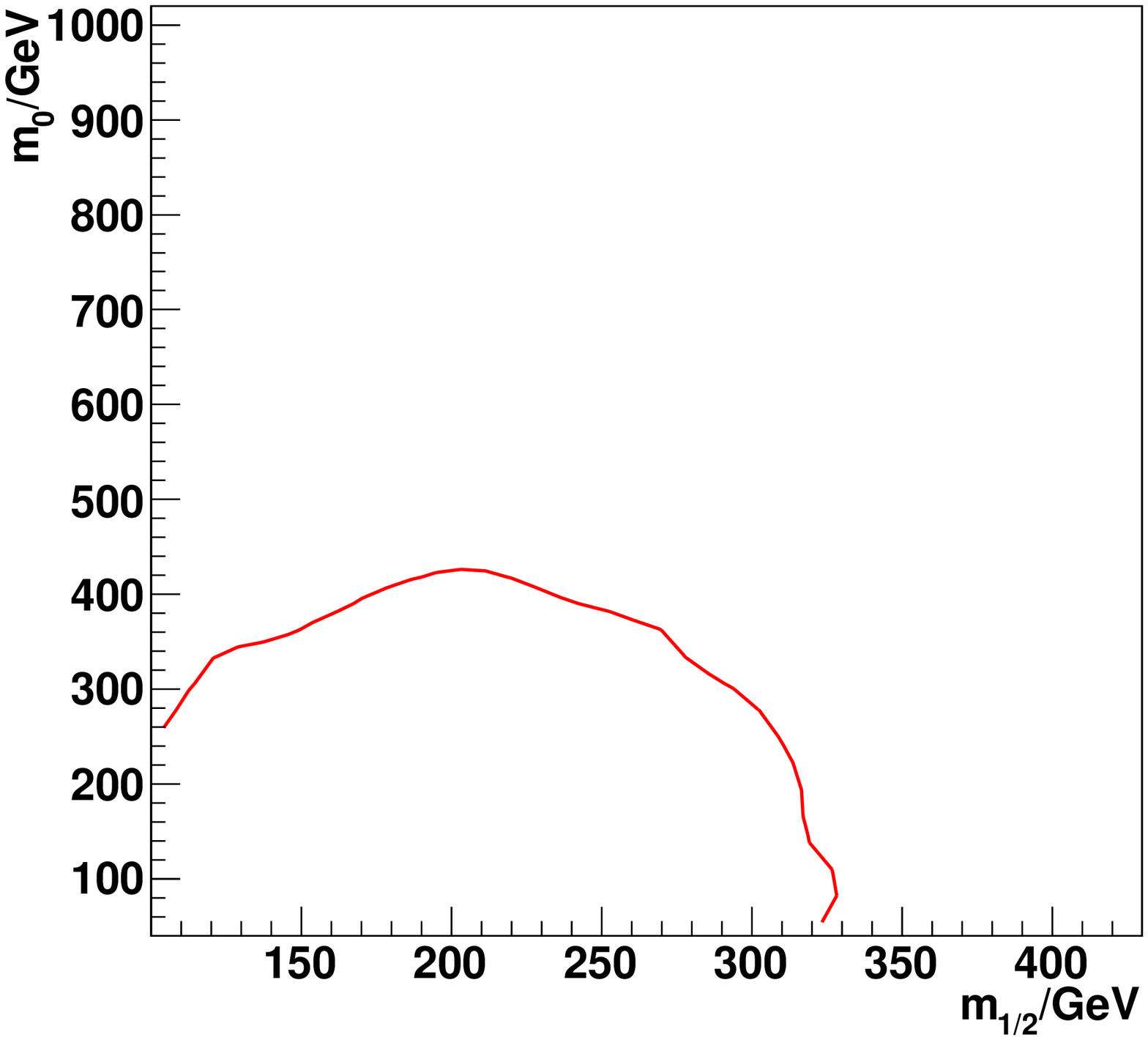}}
  \caption{The 95$\%$ exclusion contour in the $m_0,m_{1/2}$ plane of the CMSSM for $A_0=0$ and $\tan\beta=3$ as obtained from a grid of \texttt{AcerDet}-simulated points (left) and from our neural network output (right). The neural network was trained on 5000 data points, and the exclusion contour was generated by evaluating the network output for 5000 points with randomly chosen $m_0$ and $m_{1/2}$ values.}
  \label{fig:limit}
\end{figure}

To further illustrate that our neural net output adequately reproduces the behaviour of the original simulation, we show in Figure~\ref{fig:limit} the 95$\%$ exclusion contour in the CMSSM for a grid of points similar to the ATLAS grid used in Reference~\cite{daCosta:2011qk}. This features a coarse scan over $m_0$ and $m_{1/2}$, with the other parameters fixed at $A_0=0$ and $\tan\beta=3$. For each point in the grid, we assign a Poisson likelihood as above, then assign each point a value of $\Delta \chi^2 = -2\ln L/L_\text{max}$, where $L_\text{max}$ is the maximum value of the likelihood obtained in the scan. We then smooth a 2D histogram of these values and plot the contour $\Delta \chi^2=5.99$, corresponding to the 95\% exclusion contour.

The left-hand plot of Figure~\ref{fig:limit} shows the original simulation output for a grid of points in which $m_0$ and $m_{1/2}$ are chosen to reproduce the published ATLAS values. Each point in this plot takes approximately 30 minutes to process, this being dominated by the time taken to perform the event generation and detector simulation. The right-hand plot is generated using our neural net trained using 5000 data points. We first fix $A_0$ and $\tan\beta$ to the ATLAS values and then choose 5000 random points in the $m_0,m_{1/2}$ plane and evaluate the net output. Each of these takes a fraction of a second to generate. The result is a smoother limit, but one that nevertheless agrees very well with the original simulation. The bumpy sections of the left-hand plot are likely to have arisen from the poor performance of the smoothing procedure performed on the coarse grid points. The advantage of using the neural net output is that one can obtain a much finer exploration of the contour (in addition to the fact that one can explore the full space of the CMSSM rather than simply a parameter slice as in this validation example).

Comparison of Figure~\ref{fig:limit} with the original ATLAS contour in Reference~\cite{daCosta:2011qk} shows that our limit is slightly weaker. This can be understood by the fact that we have not included next-to-leading order effects in the SUSY production cross-section, which would tend to increase the yield of events per point and thus shift the limit to higher values. These could be included by weighting each of our training data points by the relevant $k$-factor. We have also not included the effect of detector systematic errors, although these could be applied in a later study by comparing our simulation output to the published ATLAS results, as was recently performed in~\cite{Allanach:2011wi}. These do not affect our proof of principle so we defer them to a future study. It is, however, worth noting that the effect of neglecting $k$-factors is to make our limit more conservative than the ATLAS contour.

\subsubsection{Results obtained using a Support Vector Machine}
The support vector regressor was trained using two kernel functions, namely the polynomial kernel and the RBF kernel.  The following three-stage process was used:
\begin{enumerate}
 \item Data pre-processing: training data was normalised to zero mean, unit variance on a per-parameter basis using ${\bf p}^{(i)} := \left( \mathrm{diag} \left( \Delta {\bf p} \right) \right)^{-1} \left( {\bf p}^{(i)} - {\bar {\bf p}} \right)$, $y^{(i)} := \left( \Delta y \right)^{-1} \left( y^{(i)} - {\bar y} \right)$ for all $i \in \mathbb{Z}_N$, where ${\bar y}$, ${\bar {\bf p}}$ and ${\Delta y}$, ${\Delta {\bf p}}$ are the means and component-wise standard deviations, respectively, of the training targets $y^{(i)}$ and input parameter vectors ${\bf p}^{(i)}$.
 \item \label{step_two} Parameter selection: the training set was split into two subsets $\Theta_T$ (training) and $\Theta_E$ (evaluation) containing, respectively, $2/3$ and $1/3$ of the total training set.  For all combinations of regression parameters $C/N \in \left\{ 0.1, 0.5, 1, 5, 10, 50, 100 \right\}$, $K \left( {\bf p}, {\bf q} \right) \in K_p \cup  K_R$:
 \begin{equation*}
  \begin{array}{rcl}
   K_p & = & \left\{ \left. \left( 1 + {\bf p}^T {\bf q} \right)^d \right| d \in \left\{ 1, 2, 3 \right\} \right\} \\
   K_R & = & \left\{ \left. \exp \left( -\frac{1}{\sigma} \left\| {\bf p} - {\bf q} \right\|^2 \right) \right| \sigma \in \left\{ 0.1, 0.5, 1, 5, 10, 50, 100 \right\} \right\} \\
  \end{array}
 \end{equation*}
 the support vector regressor was trained using $\Theta_T$ and its performance evaluated using the mean-squared prediction error on $\Theta_E$ to find the regression parameters $C/N$ and $K$ which minimised the mean-squared error.
 \item Validation: the support vector regressor was trained using the complete training set and the support vector regression parameters chosen in step \ref{step_two}.  The trained regressor was then evaluated on an independent test sample of $5000$ points.
\end{enumerate}

\begin{figure}
\begin{center}
  \subfloat[]{\label{fig:svmcomp1}\includegraphics[width=0.48\textwidth]{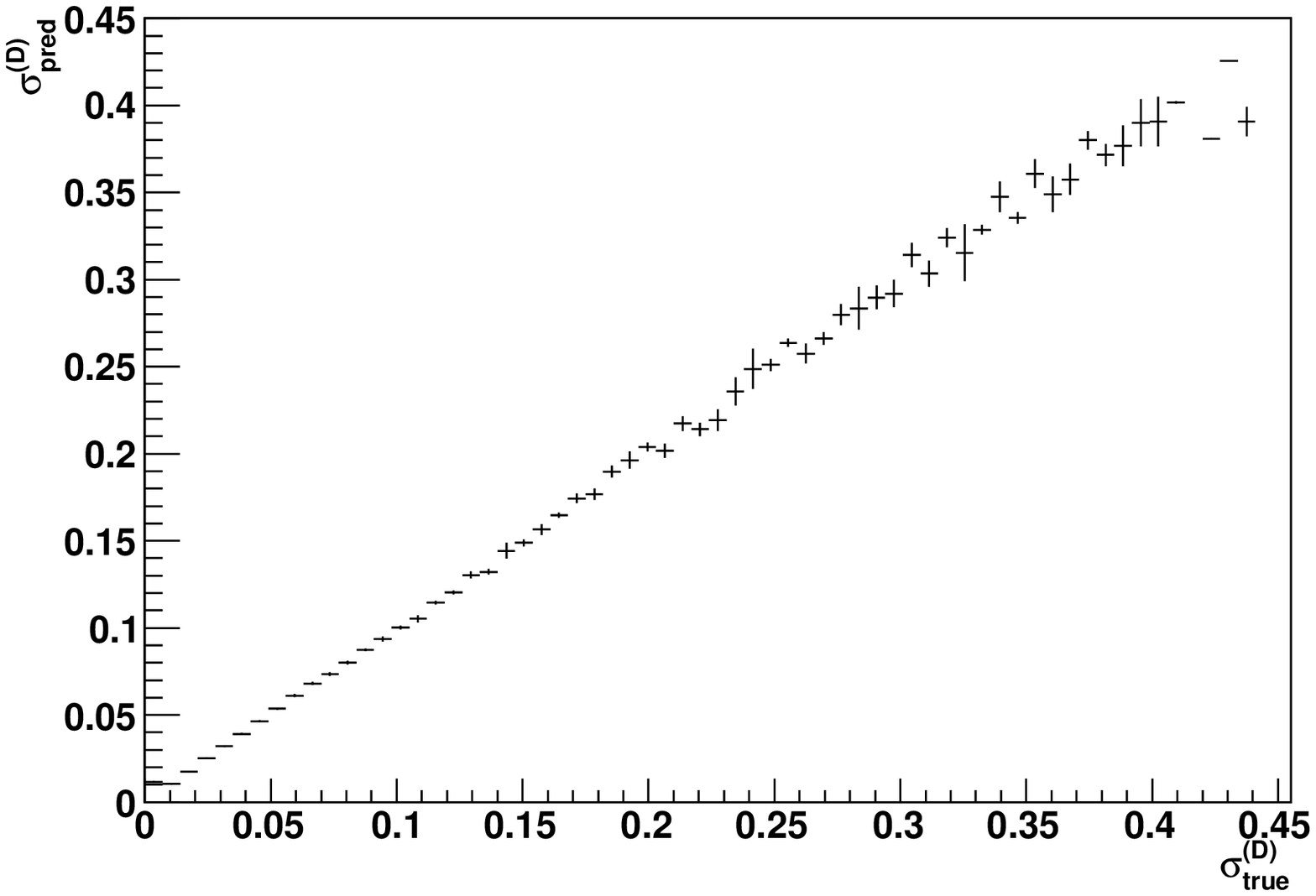}}
  \subfloat[]{\label{fig:svmcomp2}\includegraphics[width=0.48\textwidth]{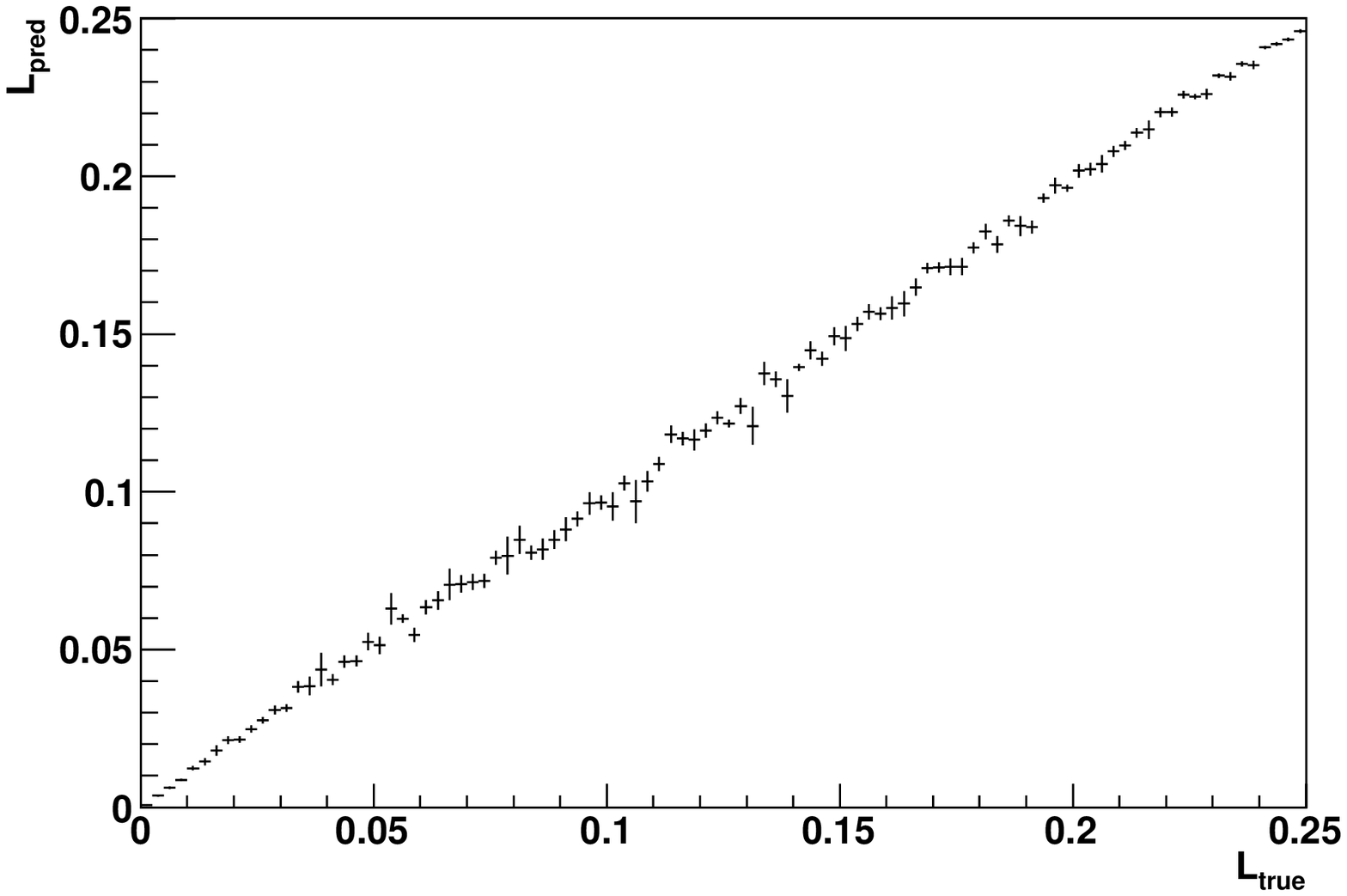}}
\caption{The true cross-section of events passing the ATLAS region D selection cuts (left), and the true Poisson likelihood of points in the CMSSM space (right) shown for the original simulation output vs. the prediction obtained using a support vector machine trained with 5000 simulated data points and evaluated on an independent test sample of 5000 points.}
\label{fig:svmcomp}
\end{center}
\end{figure}

Using this process we found regressor parameters $K \left( {\bf p}, {\bf q} \right) = \exp ( -\left\| {\bf p} - {\bf q} \right\|^2 )$ and $C/N = 5$ were optimal for a training set of $5000$ simulated data points.  Figure~\ref{fig:svmcomp1} shows the true cross-section of events in the ATLAS region D search channel vs the support vector machine output. Comparison with Figure~\ref{fig:comp1} demonstrates that the results are very similar, though the support vector machine output is better at predicting the value from the original simulation for higher cross-sections. The effect on the Poisson likelihood can be seen in Figure~\ref{fig:svmcomp2}. The advantage of the support vector machine is reduced in the likelihood since the better performance at high cross-sections will translate to only a small change in already unlikely parameter values. The SVM output looks worse at higher likelihoods, but we have checked that increasing the training set further leads to results that are on a par with the Bayesian neural net results. The obvious conclusion is that the Bayesian neural net makes a more efficient use of the training data.

\section{Variation of performance with training data}
\label{training}
\begin{figure}[p]
\sixgraphs{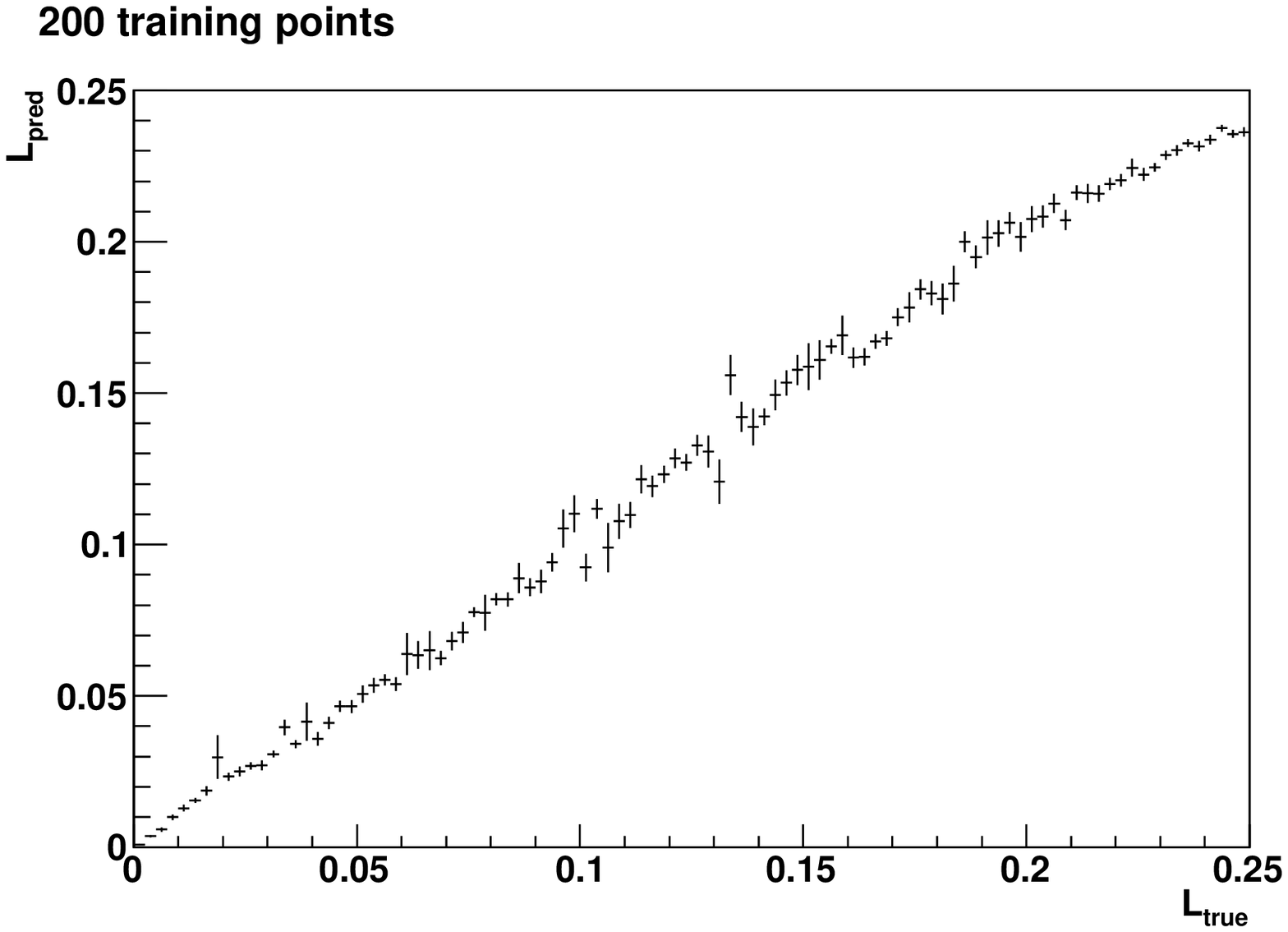}{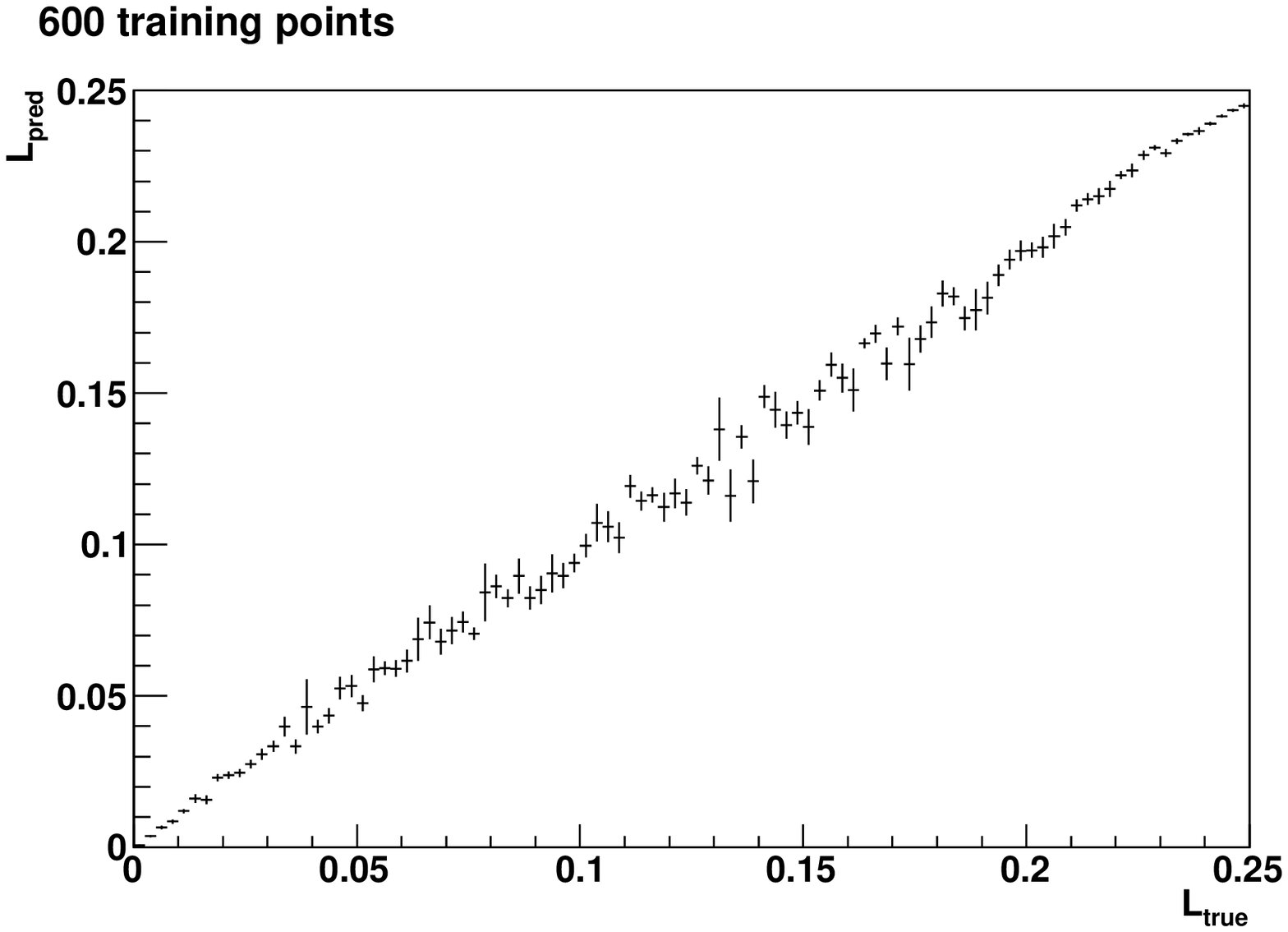}{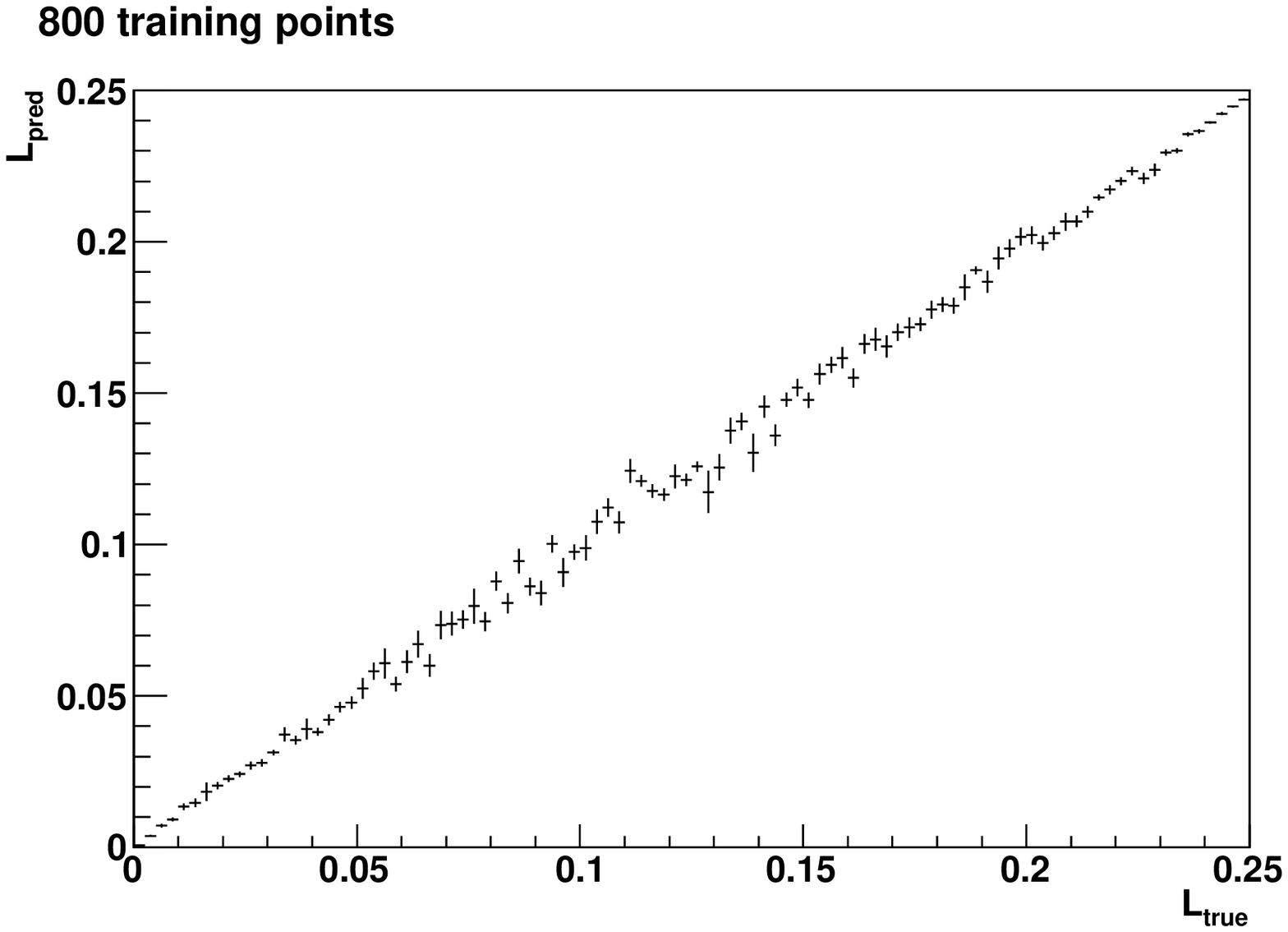}{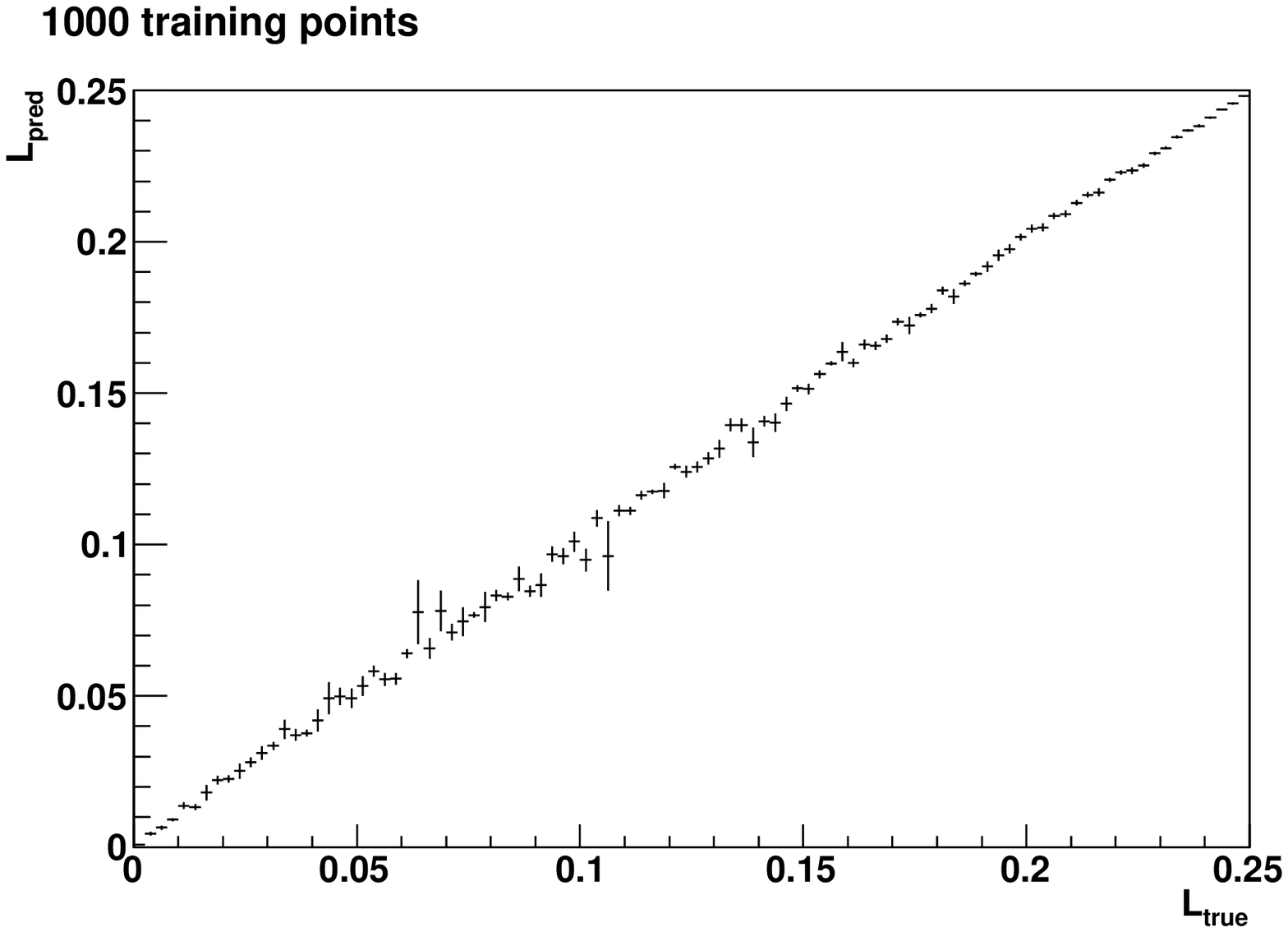}{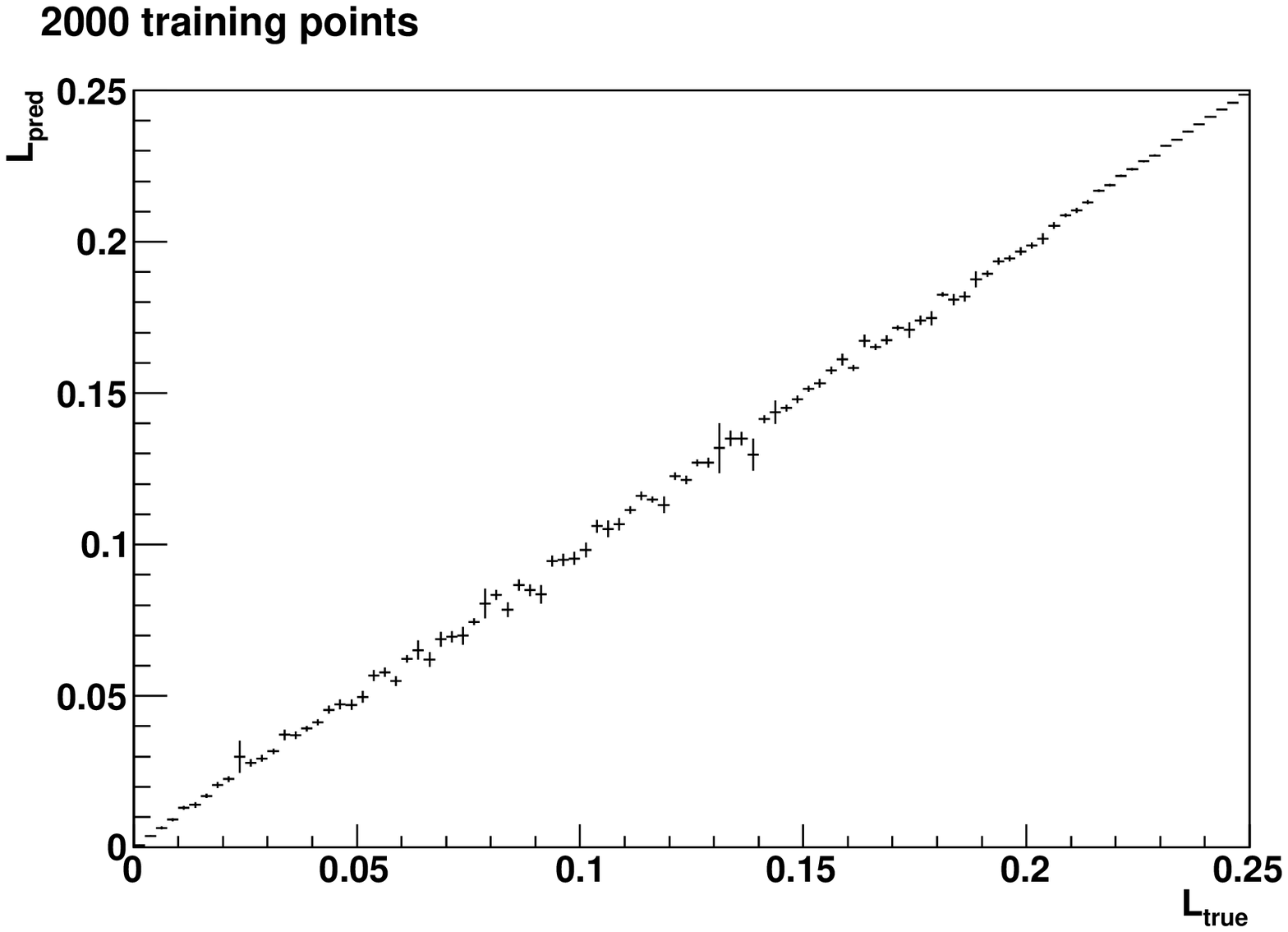}{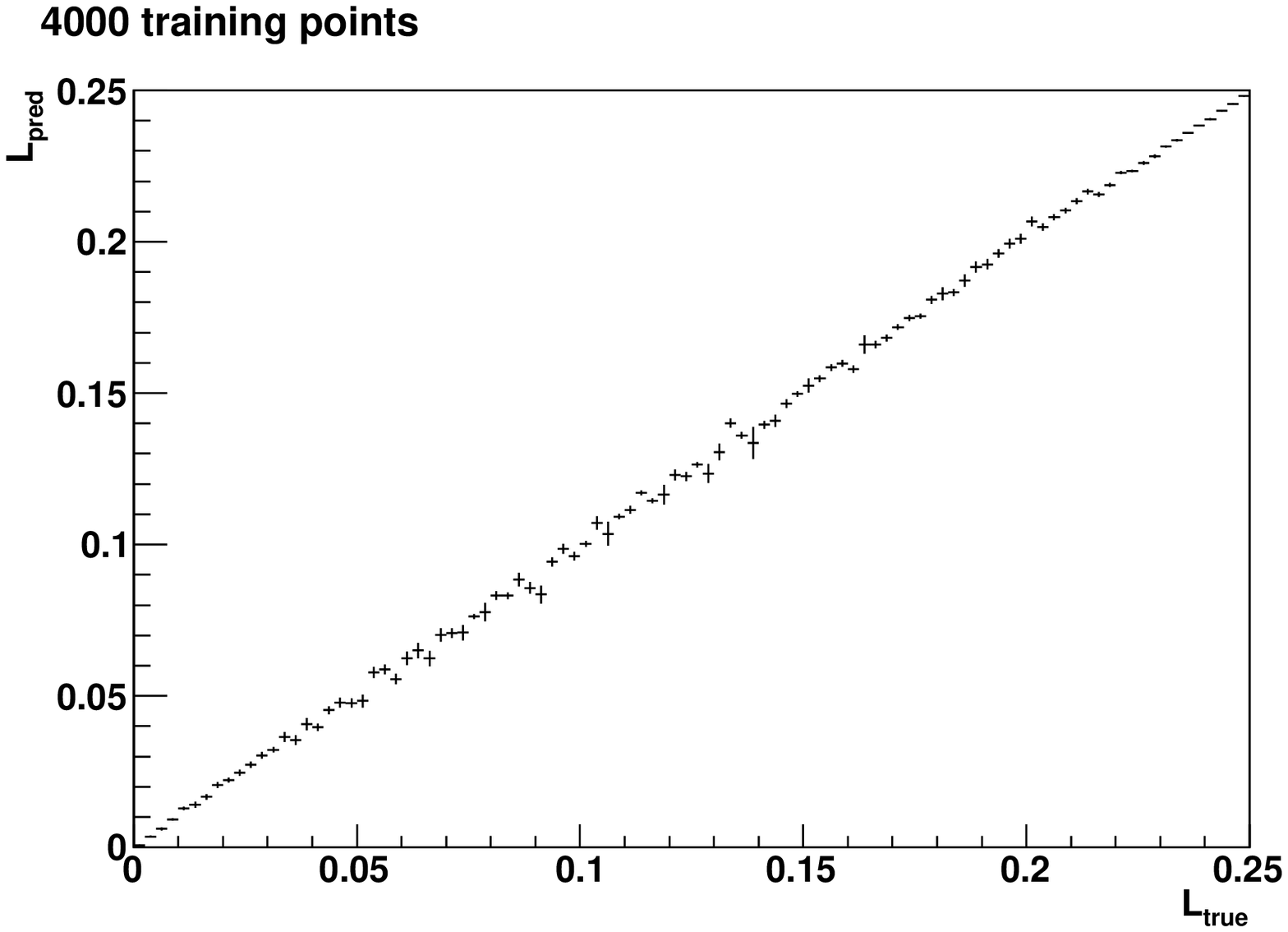}
\caption{Profile histogram of original likelihood vs Bayesian neural net output for varying amounts of training data.}
\label{like-train}
\end{figure}

Figure~\ref{like-train} shows the true likelihood value vs the Bayesian neural net prediction for different numbers of training points, evaluated on an independent test sample of 5000 points. The performance clearly degrades for low numbers of points, and gradually approaches the optimum performance obtained with 5000 training points (Figure~\ref{fig:comp2}). The number of training points required to reduce the accuracy of the likelihood prediction to the percent level would seem to be at least 2000, after which the performance increases more slowly.

\begin{figure}[p]
\sixgraphs{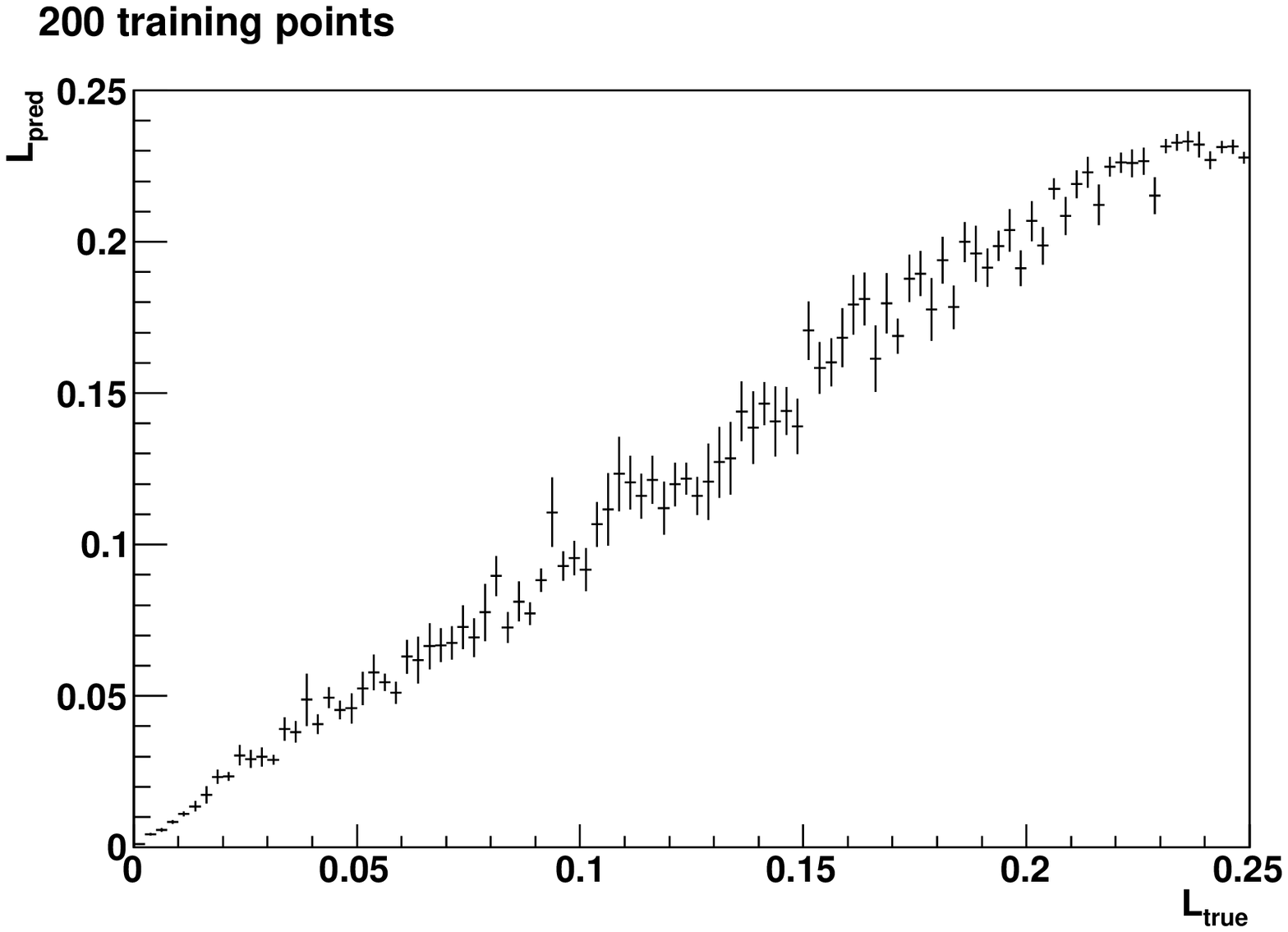}{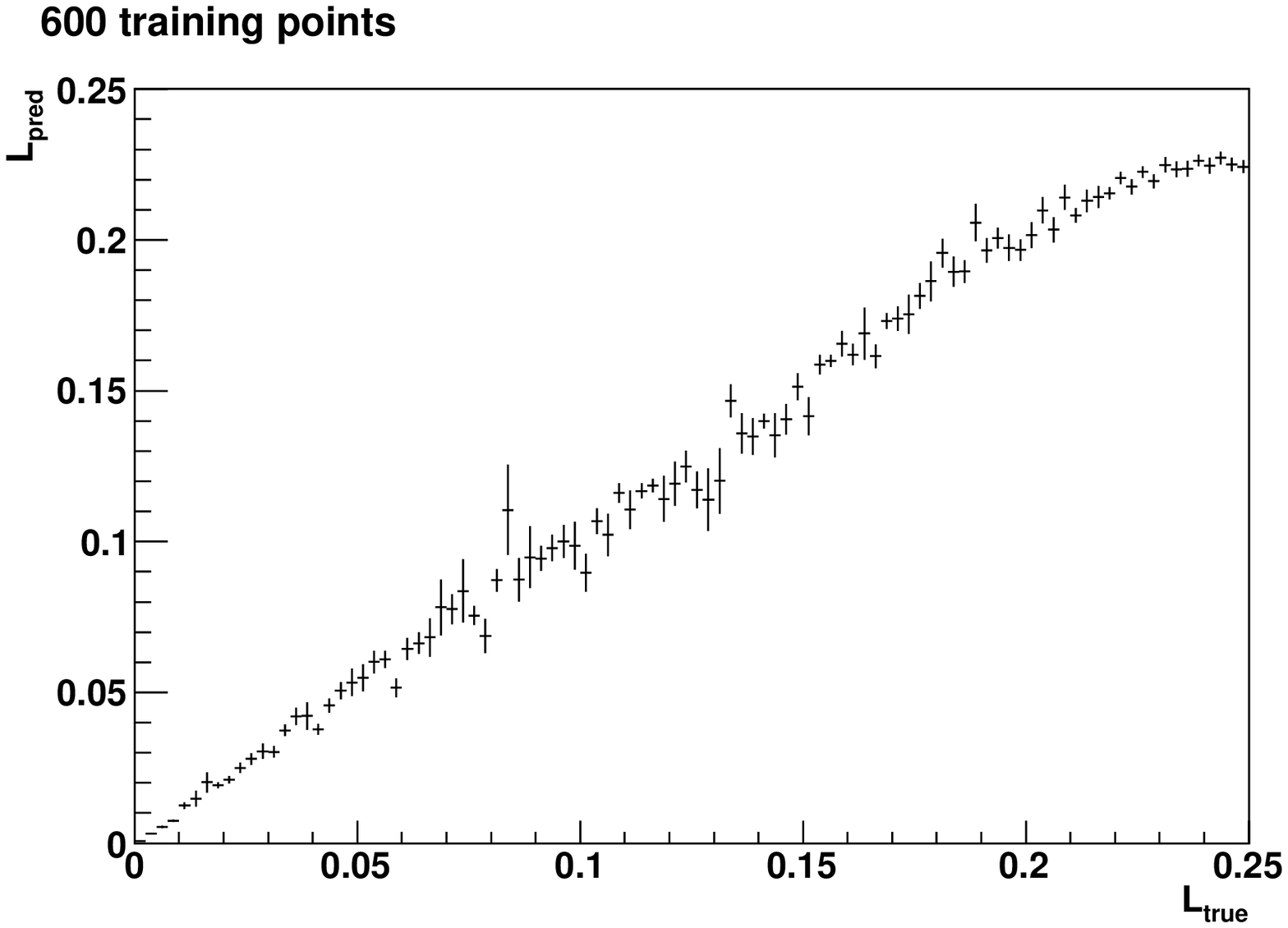}{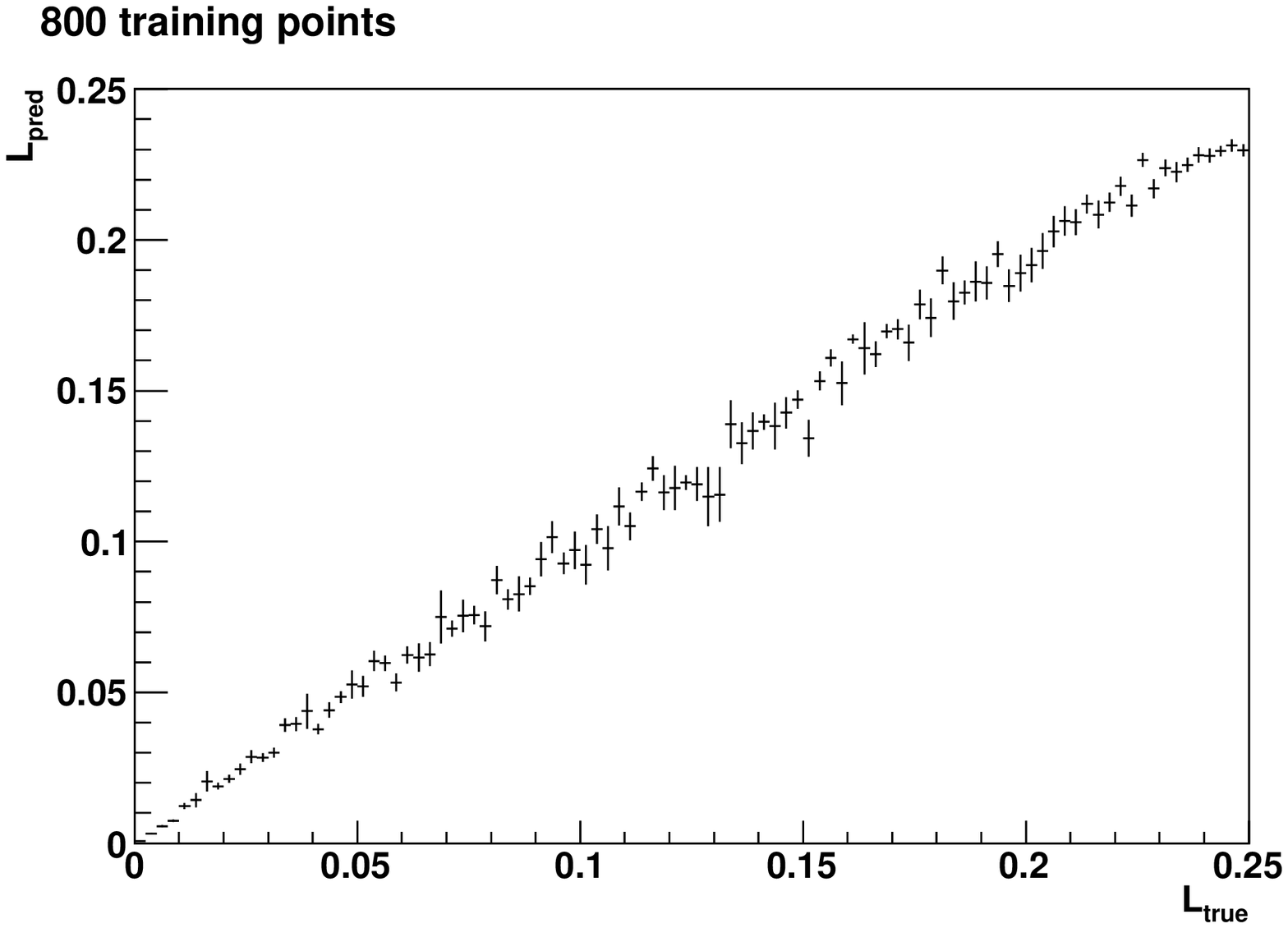}{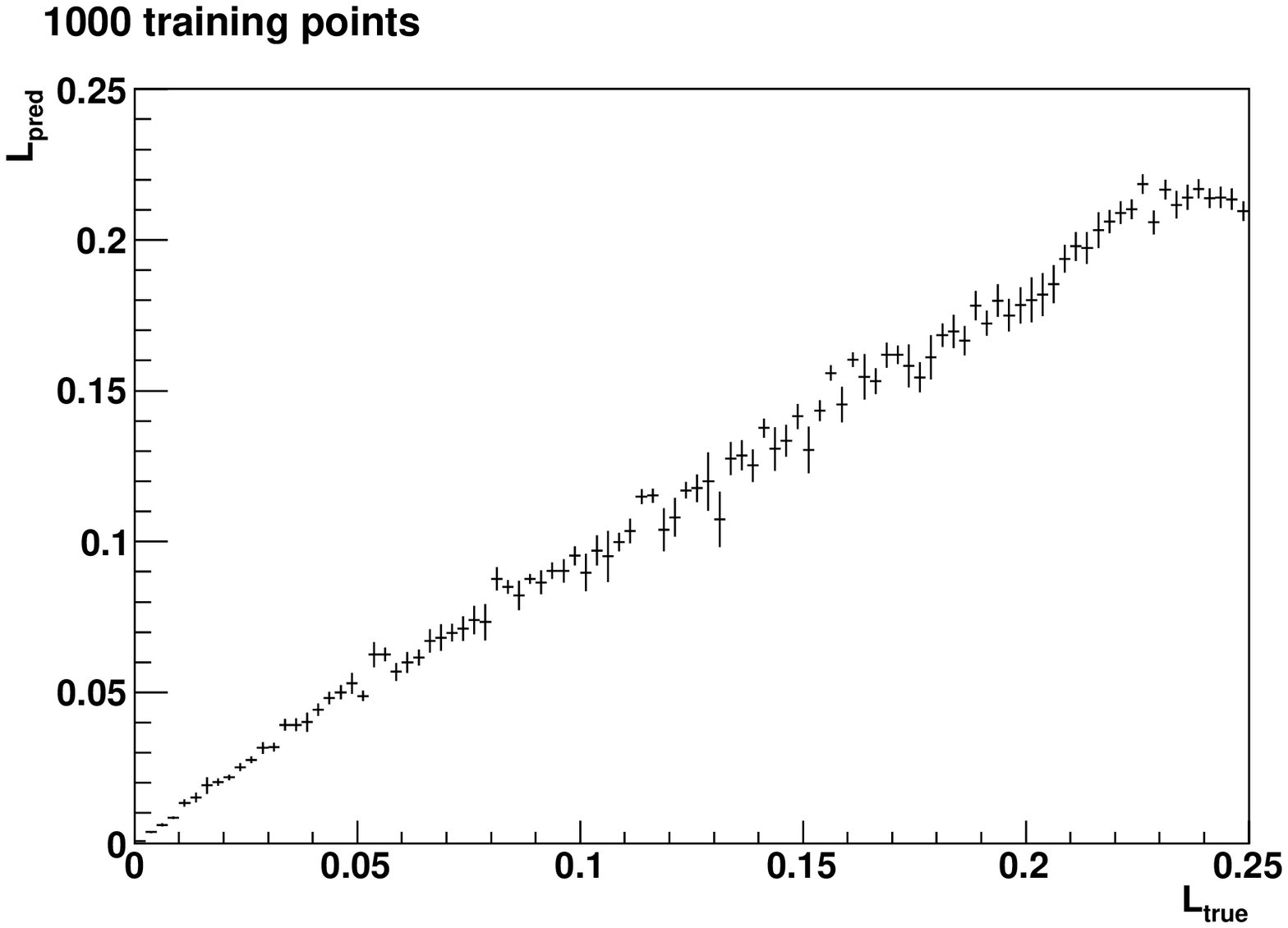}{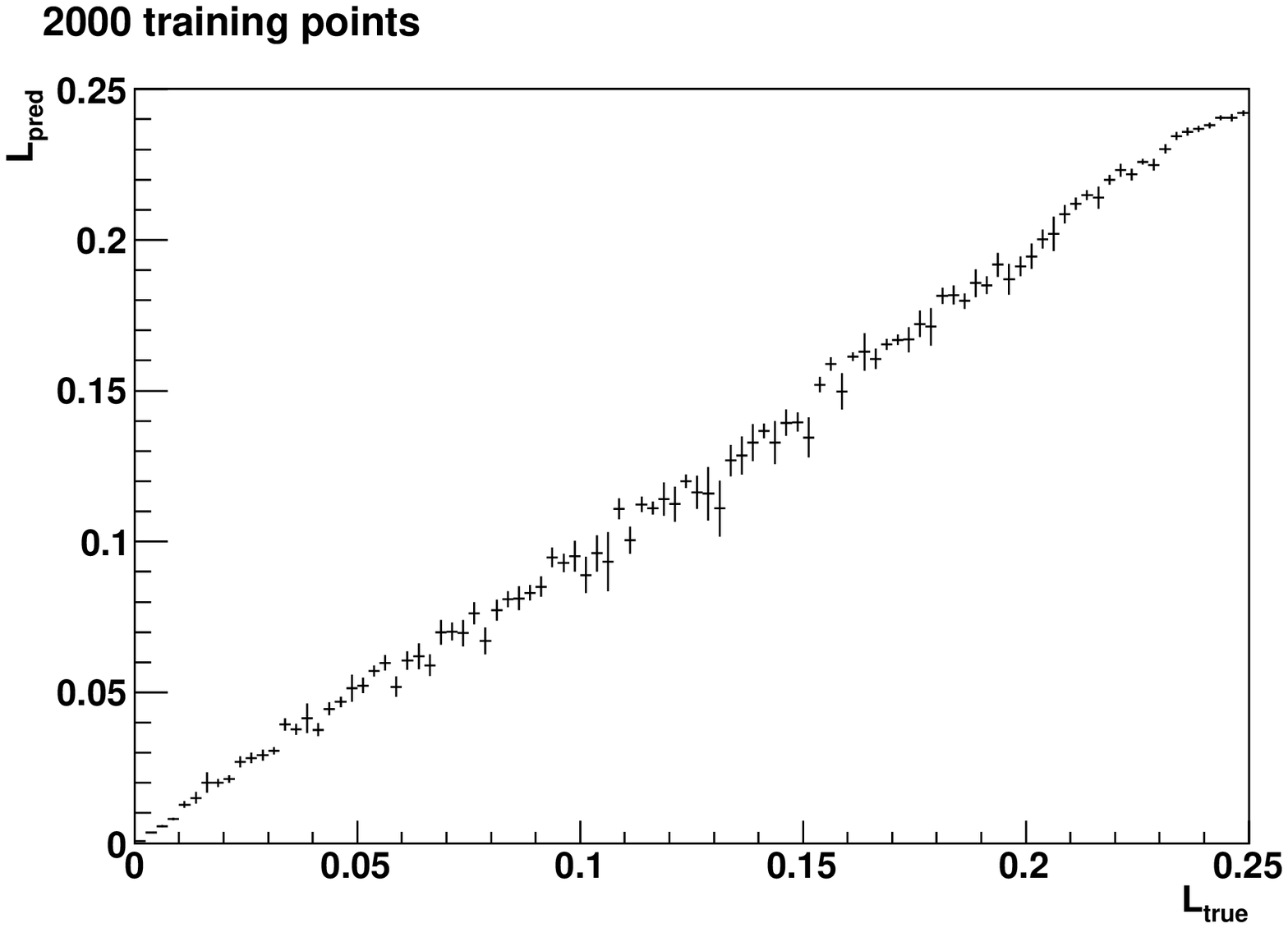}{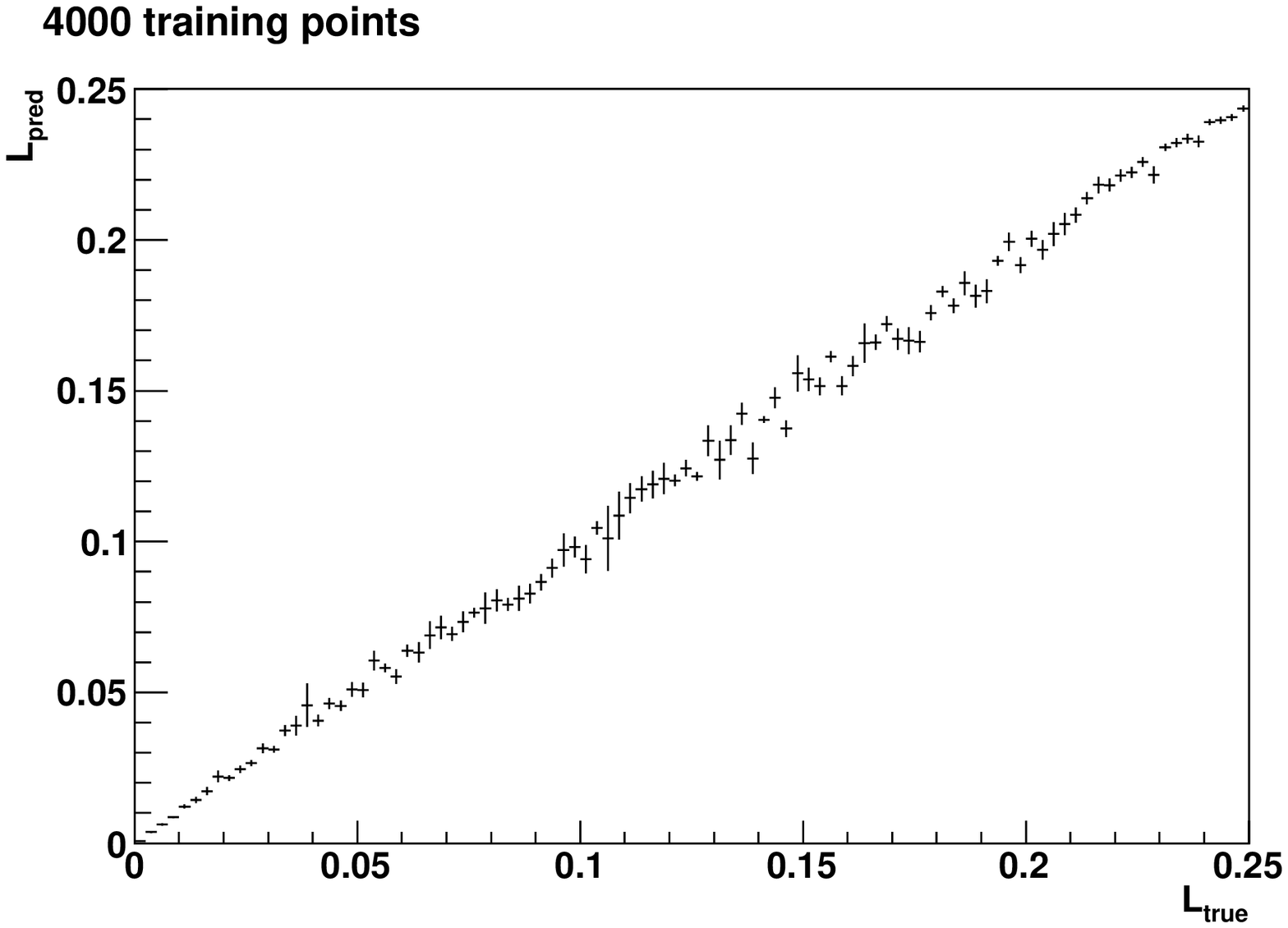}
\caption{Profile histogram of original likelihood vs support vector machine output for varying amounts of training data.}
\label{svm-train}
\end{figure}

In Figure~\ref{svm-train}, we show the equivalent results obtained using a support vector machine.  The support vector machine results are noticeably worse than the Bayesian neural net results for smaller numbers of training points, with clear evidence for both a greater spread of predicted values for each true likelihood and a systematic underestimate of the likelihood for the most likely points. The latter is particularly serious given that it is precisely the large likelihood points that one is interested in in practice. This is despite the fact that in the limit of a large amount of training data, the SVM results are competitive with the Bayesian neural net results.

\clearpage

\section{Discussion}
\label{discussion}
The results of the previous section indicate that it is indeed possible to interpolate between a grid of points in the full parameter space of the CMSSM. The training data generated within 36 hours proved easily sufficient to model the most constraining LHC search data and thus provide a quick function for phenomenological studies in the CMSSM. There should be no barrier to either repeating this with different LHC search channels, or to trying different underlying SUSY models.

Our investigation of the number of training points required to observe reasonable performance indicates that a few thousand training points are required to reduce the error in the likelihood to the per cent level when using a Bayesian neural net, and thus to observe identical output from the regressor as from the original simulation. This is ideal for phenomenological studies using fast simulations, particularly given that we managed to generate tens of thousands of training points with 50,000 events per point within 36 hours. For experimental collaborations wishing to use a full simulation, this level of simulation is still achievable given the availability of grid computing resources, but it may cease to be feasible as the integrated luminosity stored by ATLAS and CMS increases and thus the time and disk space allocated to the processing of LHC data removes the resources available for Monte Carlo simulation. Since in practice one would want to test the results on independent samples (such as those we used here), this further increases the stress on limited resources. We therefore suggest that the use of fast and semi-fast simulations would be useful tools if these can be validated against the full simulation within the collaborations.

The support vector machine results are broadly competitive with the Bayesian neural net results in the limit of a large amount of training data, but are noticeably worse for restricted training data sets. We thus conclude that phenomenologists using faster simulations may safely use either approach whilst experimental collaborations with more complex simulation requirements may prefer to use a Bayesian neural net. We further note that generation of the training data is not the only deciding factor in which approach to use. The ease of using a particular code for generating output predictions for input values not in the original sample is an important factor in designing a workable system for non-expert users, and thus having a choice of algorithms and computer codes is useful. In our study, it was indeed easier to generate output values using the support vector machine implementation.

It is useful at this point to compare the approach taken in this publication to other solutions to the problem of using LHC event rate data in SUSY parameter fits. An earlier attempt by the author and collaborators in Reference~\cite{Lester:2005je} parallelised the event generation and simulation step, thus reducing the time taken to calculate the likelihood for a given point in parameter space. This allowed a Bayesian analysis of the CMSSM to be performed on a supercomputer using a Markov Chain Monte Carlo sampling algorithm, which is a classic case of an algorithm in which (at least for a given chain), one must generate likelihoods in sequence rather than in parallel. Only 3 parameters were varied, and only 1000 events were generated per point. This has a number of disadvantages over the present method. Firstly, generating only 1000 events per point leads to large statistical fluctuations in the prediction for each point. Secondly, the parallelisation of the code was itself a major project, and the running of the code required access to a suitably fast supercomputer. Thirdly, the results were at leading order only, and adding a next-to-leading order calculation of the cross-section would lead to an unacceptably large computing time. The present method generates the training data in parallel before the parameter fit is carried out, enabling one to use a cluster which is more easily available. One can also add next-to-leading order effects or other computationally expensive calculations to the training data provided one has enough CPU resources in the cluster to generate the training data in a reasonable time, without worrying about having to calculate likelihoods in sequence.

A more recent solution to the problem of incorporating LHC event rate data into SUSY parameter fits is that given in~\cite{Dreiner:2010gv}. This uses parameterised cross-sections, branching ratios and acceptances for given search channels. Cross-sections are calculated based on the squark and gluino masses for a given point and are stored in a look-up table. Branching ratios can be evaluated very quickly using a SUSY spectrum generator. Acceptances are evaluated using a combination of look-up tables and calculations of jet and lepton energies in the squark rest frame. The approach taken may be considered similar to that taken here in the sense that one can assign a likelihood for a given point based on results generated previously. However, the parameterisation of cut acceptances must be done separately for processes with different sparticle mass hierarchies, and the simulation does not include effects such as initial and final state radiation that might make an event appear in a different search channel than that naively expected based on the SUSY decay processes that are open at that point. The present method is based on interpolation of simulation results that includes all of these effects. The results of the fast parameterisations can be made available independently of the need to generate training data, and hence can be used to build LHC event rate data into generic fitting algorithms.

Having compared the method presented here favourably to others in the literature, it is only fair to state the limitations. Firstly, one has to generate a new grid of training data if one chooses to look at a different physics model. Given the speed with which we generated our training data this would not seem a major drawback for phenomenological applications, and a similar study in different classes of model is of great interest for future study. Secondly, each search channel must be added as a target variable to the neural net separately. This itself is trivial, though one may find that, e.g., dilepton channels require a different strategy for choosing training data than the zero lepton channel used here. Thirdly, we have not incorporated the effect of systematic errors. This is acceptable if the limit we derive from our results is conservative (as it is here), but would be a major issue to address should an experimental collaboration wish to use these results. In fact, if one were to store four vector data in addition to the yield for each training point, the effect of detector systematics could be determined fairly quickly, and the effects could be added to the parameter space as nuisance parameters. This would have the added advantage that new search channels could be applied to the existing training data without having to regenerate events.
\section{Conclusions}
\label{conclusions}
We have set out to develop a method of performing fast SUSY phenomenology studies using LHC event rate data, and have presented a toy example that shows successful interpolation of simulated LHC event rate data. This information, combined with published LHC data on the number of observed, and expected background, events allows one to assign a likelihood to points in the full space of the CMSSM in fractions of a second. We find that with $\sim 2000$ training points, one can approximate the fully simulated likelihood to within a few percent across the range of interest, and our fast simulation results of the ATLAS zero lepton search already approximate the published ATLAS exclusion limit. This proof of principle presents a very promising avenue for future work, and promises a general and intuitive approach for using LHC data in phenomenological studies.

\section{Acknowledgements}
We are greatly indebted to Mike Hobson for many interesting and useful conversations at the OKC Prospects Workshop, Stockholm in September 2010. We also owe thanks to Tim Dyce of the University of Melbourne for assistance in using the Melbourne Tier 2 cluster, and to the Melbourne members of the Australian Research Council (ARC) Centre of Excellence for Particle Physics at the Terascale for useful discussions during preparation of the final draft. MJW is supported by ARC Discovery Project DP1095099.  AS is supported by the ARC Research Network on Intelligent Sensors, Sensor Networks and Information Processing (ISSNIP). AB is supported by a Scottish Universities Physics Alliance research fellowship.
\bibliography{bibfile,universal}
\end{document}